\documentclass[twocolumn,floatfix,nofootinbib]{revtex4}
\usepackage{graphicx}
\usepackage{amsmath}
\usepackage{amssymb}
\usepackage{bm}

\begin{document}

\title{Wigner-Poisson statistics of topological transitions in a Josephson junction}
\author{C. W. J. Beenakker}
\affiliation{Instituut-Lorentz, Universiteit Leiden, P.O. Box 9506, 2300 RA Leiden, The Netherlands}
\author{J. M. Edge}
\affiliation{Instituut-Lorentz, Universiteit Leiden, P.O. Box 9506, 2300 RA Leiden, The Netherlands}
\author{J. P. Dahlhaus}
\affiliation{Instituut-Lorentz, Universiteit Leiden, P.O. Box 9506, 2300 RA Leiden, The Netherlands}
\author{D. I. Pikulin}
\affiliation{Instituut-Lorentz, Universiteit Leiden, P.O. Box 9506, 2300 RA Leiden, The Netherlands}
\author{Shuo Mi}
\affiliation{Instituut-Lorentz, Universiteit Leiden, P.O. Box 9506, 2300 RA Leiden, The Netherlands}
\author{M. Wimmer}
\affiliation{Instituut-Lorentz, Universiteit Leiden, P.O. Box 9506, 2300 RA Leiden, The Netherlands}

\date{May 2013}
\begin{abstract}
The phase-dependent bound states (Andreev levels) of a Josephson junction can cross at the Fermi level, if the superconducting ground state switches between even and odd fermion parity. The level crossing is topologically protected, in the absence of time-reversal and spin-rotation symmetry, irrespective of whether the superconductor itself is topologically trivial or not. We develop a statistical theory of these topological transitions in an $N$-mode quantum-dot Josephson junction, by associating the Andreev level crossings with the real eigenvalues of a random non-Hermitian matrix. The number of topological transitions in a $2\pi$ phase interval scales as $\sqrt{N}$ and their spacing distribution is a hybrid of the Wigner and Poisson distributions of random-matrix theory.
\end{abstract}
\maketitle

\begin{figure}[tb]
\centerline{\includegraphics[width=0.8\linewidth]{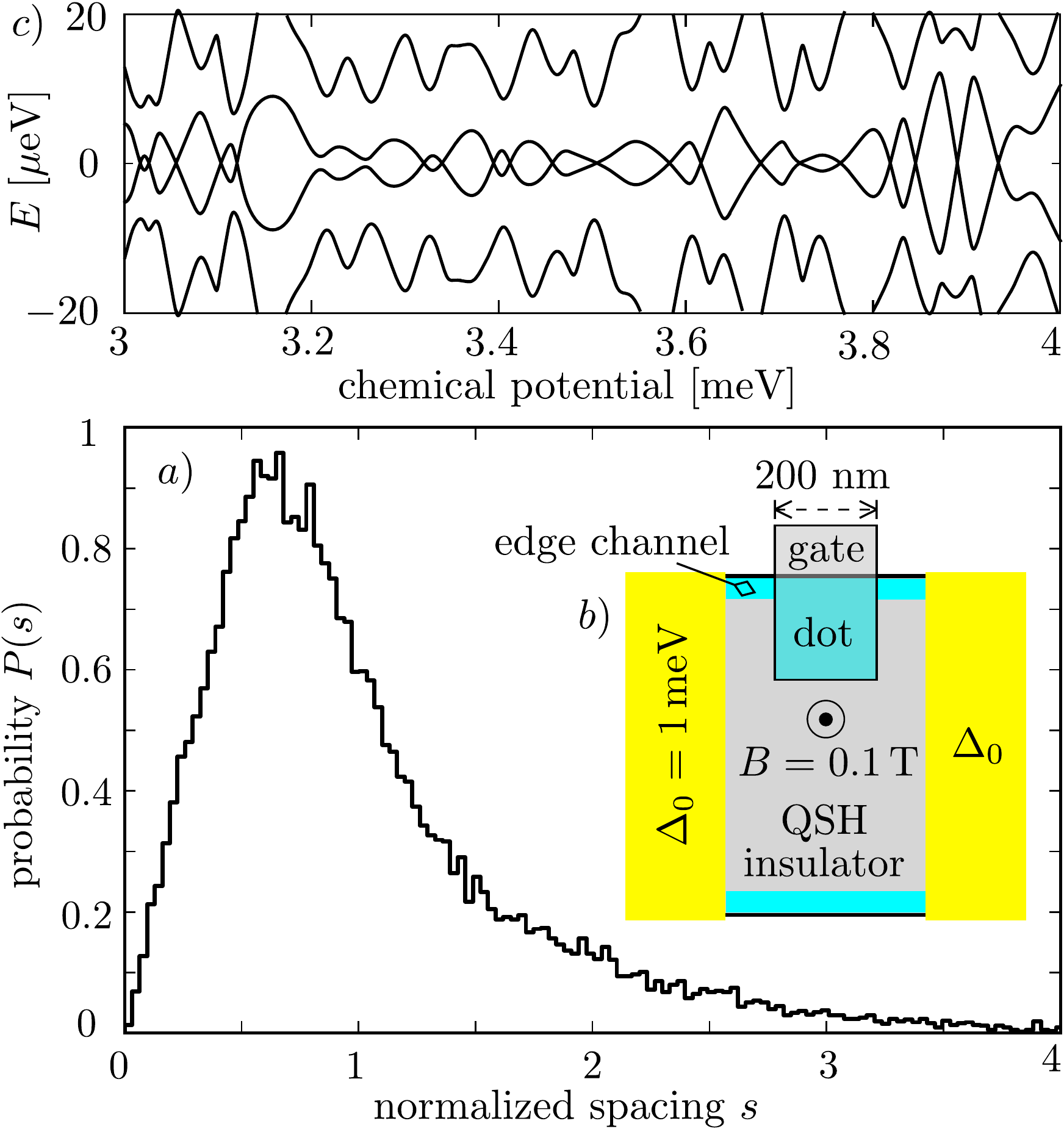}}
\caption{Model calculation of level crossings for a quantum-dot Josephson junction in an InAs-GaSb quantum well (material parameters as in Ref.\ \onlinecite{Mi13}). Panel \textit{a} shows the spacing distribution, sampled over disorder realizations, for $\simeq 50$ level crossings in a 3--6~meV chemical potential interval. Panel \textit{b} shows the geometry of the device, panel \textit{c} shows the level crossings for a single sample. 
}
\label{fig_dot}
\end{figure}

The von Neumann-Wigner theorem of quantum mechanics forbids the crossing of two energy levels when some parameter is varied, unless the corresponding wave functions have a different symmetry \cite{Neu29}. One speaks of \textit{level repulsion}. In disordered systems, typical for condensed matter, one would not expect any symmetry to survive and therefore no level crossing to appear. This is indeed the case in normal metals --- but not in superconductors, where level crossings at the Fermi energy are allowed \cite{Alt97}. The symmetry that protects the level crossing is called \textit{fermion parity} \cite{Kit01}: The parity of the number of electrons in the superconducting condensate switches between even and odd at a level crossing. To couple the two levels and open up a gap at the Fermi level one would need to add or remove an electron from the condensate, which is forbidden in a closed system.

Fermion-parity switches in superconductors have been known since the 1970's \cite{Sak70,Bal06}, but recently they have come under intense investigation in connection with Majorana fermions and topological superconductivity \cite{Ryu10,And11,Law11,Bee13,Yok13,Lee12,Cha12,Sau12,Ila13}. A pair of Majorana zero-modes appears at each level crossing and the absence of level repulsion expresses the fact that two Majorana fermions represent one single state \cite{Alt97}. Topologically nontrivial superconductors are characterized by an odd number of level crossings when the superconducting phase is advanced by $2\pi$, resulting in a $4\pi$-periodicity of the Josephson effect \cite{Kit01,Kwo04}.

Here we announce and explain an unexpected discovery: Sequences of fermion-parity switches are not independent. As illustrated in Fig.\ \ref{fig_dot}, for a quantum dot model Hamiltonian \cite{Mi13}, the level crossings show an \textit{antibunching} effect, with a spacing distribution that vanishes at small spacings. This is reminiscent of level repulsion, but we find that the spacing distribution is distinct from the Wigner distribution of random Hamiltonians \cite{Meh04,For10}. Instead, it is a hybrid between the Wigner distribution (linear repulsion) for small spacings and the Poisson distribution (exponential tail) for large spacings. A hybrid Wigner-Poisson (= ``mermaid'') statistics has appeared once before in condensed matter physics, at the Anderson metal-insulator transition \cite{Shk93,Bog99}. We construct an ensemble of non-Hermitian matrices that describes the hybrid statistics, in excellent agreement with simulations of a microscopic model.

\begin{figure}[tb]
\centerline{\includegraphics[width=0.8\linewidth]{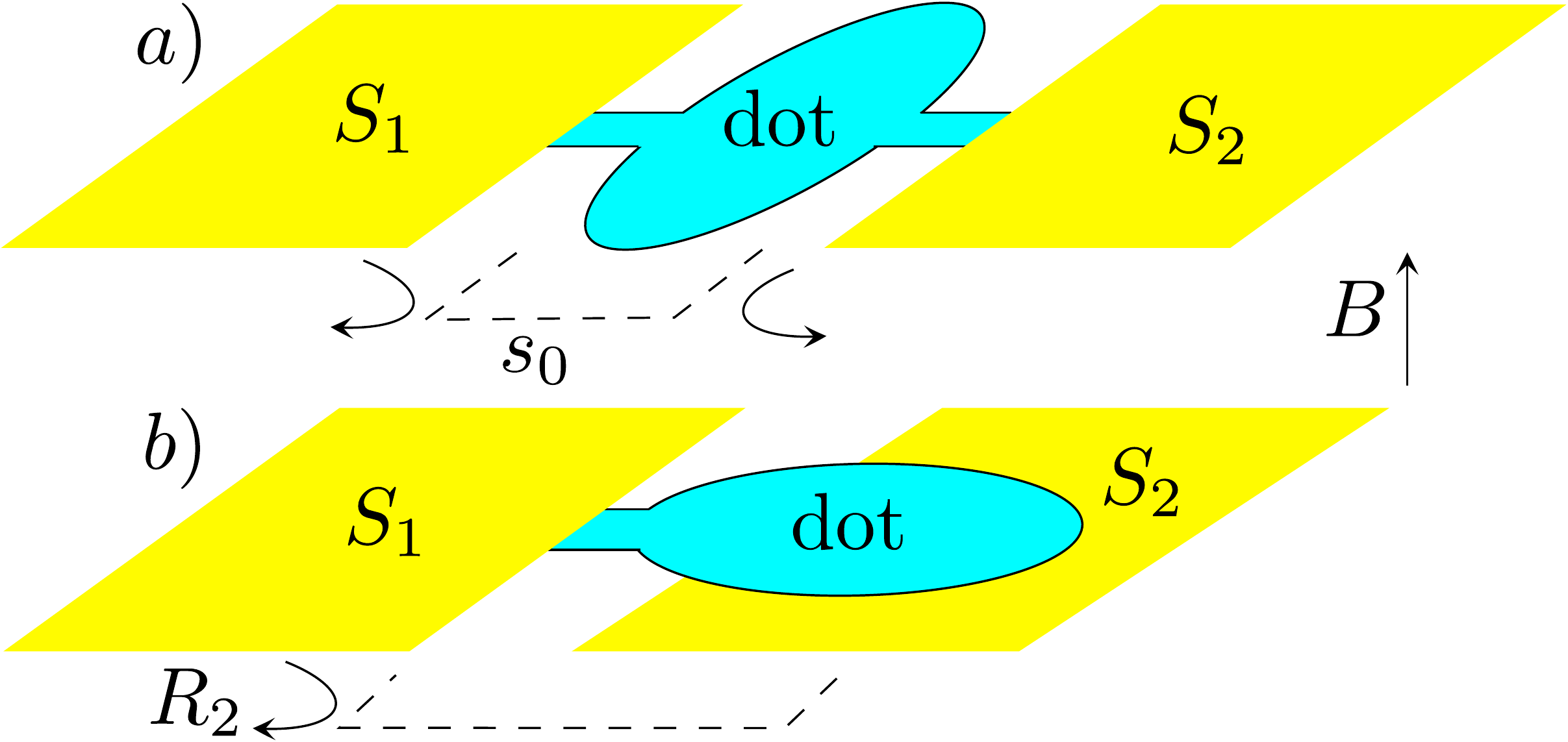}}
\caption{Two designs of a quantum-dot Josephson junction. In \textit{a} the quantum dot is coupled locally by point contacts to both superconductors $S_1$ and $S_2$, while in \textit{b} the coupling to $S_1$ is local while $S_2$ is coupled uniformly to the entire phase space of the quantum dot. In \textit{a} the chaotic scattering refers only to the normal-state scattering matrix $s_{0}$, while in \textit{b} the combined reflection from dot plus $S_{2}$ is described by a chaotic scattering matrix $R_{2}$.
}
\label{fig_setup}
\end{figure}

The geometry considered is shown in Fig.\ \ref{fig_setup}. It is an Andreev billiard \cite{Bee05}, a semiconductor quantum dot with chaotic potential scattering and Andreev reflection at superconductors $S_1$ and $S_2$. We distinguish two types of coupling to the superconductors: a strong local coupling by a ballistic point contact and a weak uniform coupling by a tunnel barrier. In Fig.\ \ref{fig_setup}\textit{a} both superconductors are coupled by a ballistic point contact with $N$ propagating modes (counting spin). The chaotic scattering in the quantum dot (mean level spacing $\delta$) then does not mix electrons and holes, on the time scale $\tau_{\rm A}\simeq\hbar/N\delta$ between Andreev reflections at the point contacts. In Fig.\ \ref{fig_setup}\textit{b} only $S_{1}$ is coupled locally. The uniform coupling to the other superconductor $S_{2}$ ensures that the entire phase space of electrons and holes is mixed chaotically within a time $\tau_{\rm A}$. These two geometries correspond to different random-matrix ensembles, essentially two extreme cases, but we will see that the statistical results are very similar.

We need to break both spin-rotation and time-reversal symmetry (symmetry class D), in order to protect the level crossings \cite{Alt97}. Spin-rotation symmetry is broken by spin-orbit coupling on a time small compared to $\tau_{\rm A}$. Time-reversal symmetry is broken by a  magnetic field $B$. A weak field is sufficient, one flux quantum $h/e$ through the quantum dot and negligible Zeeman energy, so we may assume that the spin-singlet \textit{s}-wave pairing in $S_{n}$ remains unperturbed. One then has a topologically trivial superconductor in symmetry class D, without the Majorana fermions associated with a band inversion \cite{Ali12}. 

We choose a gauge where the order parameter $\Delta_{0}$ in $S_{2}$ is real, while $S_{1}$ is phase biased at $e^{i\phi}\Delta_{0}$. The excitation spectrum of this Josephson junction is discrete for $|E|<\Delta_{0}$ and $\pm E$ symmetric because of electron-hole symmetry. As $\phi$ is advanced by $2\pi$, pairs of excitation energies $\pm E_{n}(\phi)$ may cross. The associated $\mathbb{Z}_{2}$ topological quantum number switches between $\pm 1$ at each level crossing \cite{Kit01}, indicating a switch between even and odd number of electrons in the ground state. At a constant total electron number, the switch in the ground-state fermion parity is accompanied by the filling or emptying of an excited state. We seek the statistics of these topological transitions.

The geometry of Fig.\ \ref{fig_setup}\textit{b} is somewhat easier to analyze than \ref{fig_setup}\textit{a}, so we do that first. Electrons and holes ($e,h$) at the Fermi level propagate through the point contact between $S_{1}$ and the quantum dot in one of the $N=2M$ modes. (The factor of two accounts for the $\uparrow,\downarrow$ spin degree of freedom.) Left-moving quasiparticles are Andreev reflected by $S_{1}$ and right-moving quasiparticles are reflected by the quantum dot coupled to $S_{2}$. The vector $\Psi=(\Psi_{e\uparrow},\Psi_{e\downarrow},\Psi_{h\uparrow},\Psi_{h\downarrow})$ of wave amplitudes is transformed as $\Psi\mapsto R_{2}R_{1}\Psi$, by multiplication with the reflection matrices
\begin{align}
&R_{1}(\phi)=\begin{pmatrix}
0&e^{-i\phi}\Lambda\\
e^{i\phi}\Lambda^{\rm T}&0
\end{pmatrix},\;\;\Lambda=\bigoplus_{m=1}^{M}\begin{pmatrix}
0&-i\\
i&0
\end{pmatrix},\label{RAdef}\\
&R_{2}=\begin{pmatrix}
r_{ee}&r_{eh}\\
r_{he}&r_{hh}
\end{pmatrix},\;\;r_{hh}=r_{ee}^{\ast},\;\;r_{eh}=r_{he}^{\ast}.\label{Rdef}
\end{align}
These are $2N\times 2N$ unitary matrices, with four $N\times N$ subblocks related by electron-hole symmetry. The sign of the determinant of the reflection matrix distinguishes topologically trivial from nontrivial superconductivity \cite{Akh11}. We take both $S_{1}$ and $S_{2}$ trivial by fixing ${\rm Det}\,R_{n}=1$. (The topologically nontrivial case is considered later on.)

The condition for a level crossing at phase $\phi$ is that $\Psi$ is an eigenstate of $R_{2}R_{1}(\phi)$ with unit eigenvalue, so
\begin{equation}
{\rm Det}\,[1-R_{2}R_{1}(\phi)]=0.\label{DetRphi}
\end{equation}
We seek to rewrite this as an eigenvalue equation for some real matrix ${\cal M}$. For that purpose we change variables from phase $\phi\in(-\pi,\pi)$ to quasienergy $\varepsilon=\tan(\phi/2)\in(-\infty,\infty)$. Eq.\ \eqref{DetRphi} then takes the form
\begin{equation}
{\rm Det}\,[1-U-i\varepsilon(1+U)\tau_{z}]=0,\;\;\tau_{z}=\begin{pmatrix}
1&0\\
0&-1
\end{pmatrix},\label{DetReps}
\end{equation}
with $U=R_{2}R_{1}(0)$. The Pauli matrix $\tau_{z}$ acts on the electron-hole blocks, to be distinguished from the Pauli spin matrix $\sigma_{z}$.

The complex unitary matrix $U$ becomes a real orthogonal matrix $O$ upon a change of basis,
\begin{equation}
O=\Omega^{\dagger}U\Omega,\;\;\Omega=\frac{1}{\sqrt{2}}\begin{pmatrix}
1&i\\
1&-i
\end{pmatrix}.\label{ORU}
\end{equation}
Note that ${\rm Det}\,O={\rm Det}\,U=1$, so $O\in{\rm SO}(2N)$ is special orthogonal. Since $\Omega^{\dagger}\tau_{z}\Omega=-\tau_{y}$, the level crossing condition becomes
\begin{equation}
{\rm Det}\,[1-O+\varepsilon (1+O)J]=0,\;\;J=i\tau_{y}=\begin{pmatrix}
0&1\\
-1&0
\end{pmatrix}.\label{DetO}
\end{equation}
For chaotic scattering $O$ is uniformly distributed with the Haar measure of ${\rm SO}(2N)$. This is the circular real ensemble (CRE) of random-matrix theory in symmetry class D \cite{Alt97,Bee11}.

The special orthogonal matrix $O$ can be represented by an antisymmetric real matrix $A=-A^{\rm T}$, through the Cayley transform \cite{note1}
\begin{equation}
O=(1-A)(1+A)^{-1}.\label{AOrelation}
\end{equation}
Substitution of Eq.\ \eqref{AOrelation} into Eq.\ \eqref{DetO} gives the level crossing condition as an eigenvalue equation,
\begin{equation}
{\rm Det}\,({\cal M}-\varepsilon)=0,\;\;{\cal M}=AJ=(1-O)(1+O)^{-1}J.\label{DetA}
\end{equation}
The matrix ${\cal M}$ is real but not symmetric: ${\cal M}^{\rm T}=-J{\cal M}J$. This is the definition of a \textit{skew-Hamiltonian} matrix. There are $N$ distinct eigenvalues, each with multiplicity two \cite{note2}. The $N_{X}$ distinct real eigenvalues $\varepsilon_{n}$ identify the level crossings at $\phi_{n}=2\,{\rm arctan}\,\varepsilon_{n}$.

\begin{figure}[tb]
\centerline{\includegraphics[width=0.8\linewidth]{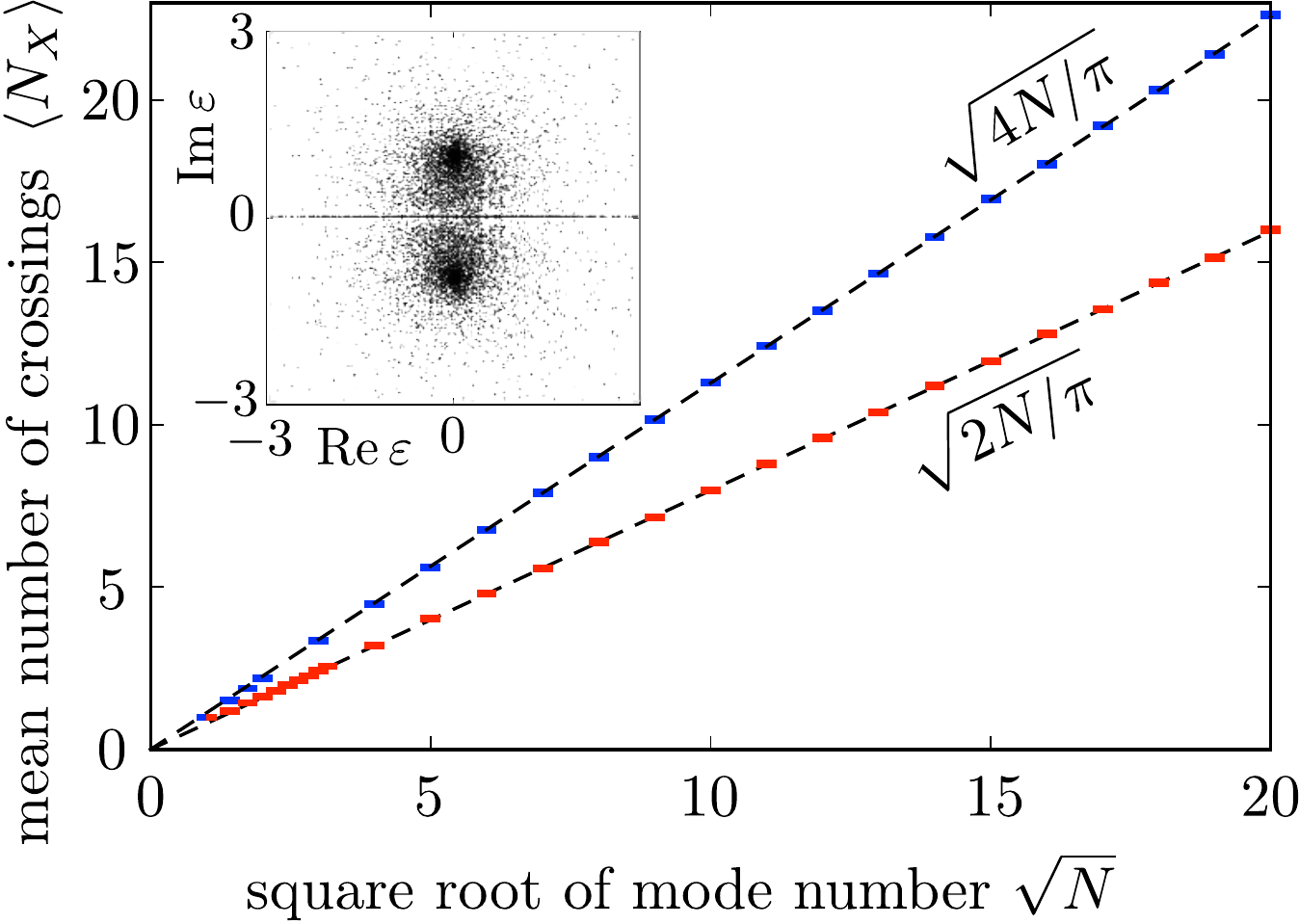}}
\caption{Plot of the $N$-dependence of the average number $\langle N_{X}\rangle$ of distinct real eigenvalues $\varepsilon$, calculated for the skew-Hamiltonian ensemble constructed from a uniformly distributed $s_{0}\in{\rm U}(2N)$ (red data points, with a scatter plot for $N=50$ in the inset) and $O\in{\rm SO}(2N)$ (blue data points). These are the expected number of level crossings (topological transitions) in a $2\pi$ phase interval for the quantum-dot Josephson junction in Fig.\ \ref{fig_setup}\textit{a} (red) and \ref{fig_setup}\textit{b} (blue). The analytical formulas given by the dashed lines have the status of a conjecture.
}
\label{fig_squareN}
\end{figure}

We have thus transformed the level crossing problem into a classic problem of random-matrix theory \cite{Leh91,Ede94,Kan05,For07,Kho11}: \textit{How many eigenvalues of a real matrix are real?} One might have guessed that an eigenvalue is exactly real with vanishing probability, since the real axis has measure zero in the complex plane. Instead, the eigenvalues of real non-Hermitian matrices accumulate on the real axis (see Fig.\ \ref{fig_squareN}, inset). This accumulation is a consequence of the fact that the complex eigenvalues come in pairs $\varepsilon,\varepsilon^{\ast}$, so real eigenvalues are stable: They cannot be pushed into the complex plane by a weak perturbation. 

\begin{figure}[tb]
\centerline{\includegraphics[width=0.8\linewidth]{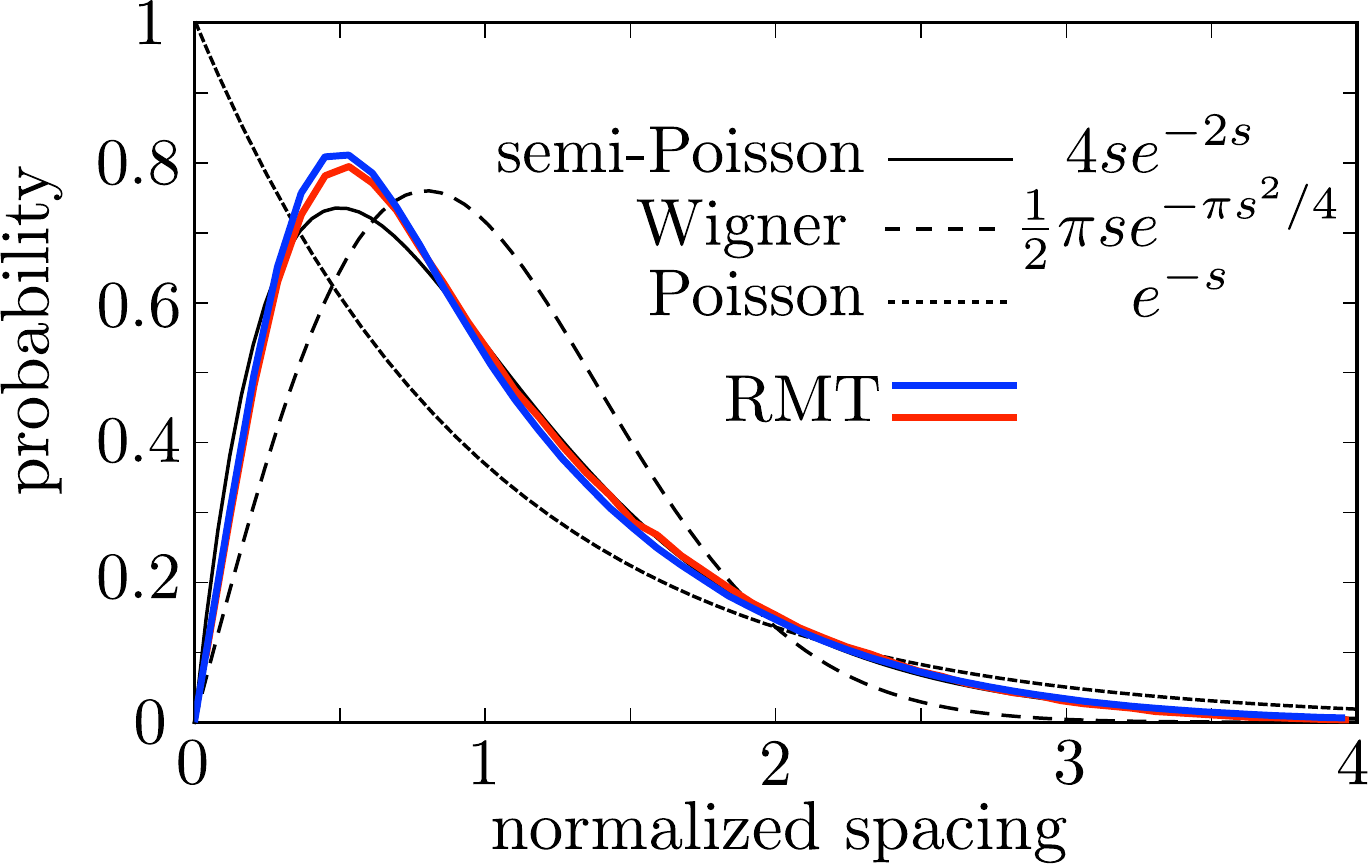}}
\caption{Probability distribution of the normalized spacings $s=\delta\phi/\langle\delta\phi\rangle$ of level crossings, calculated for the random-matrix ensemble of Fig.\ \ref{fig_setup}\textit{a} (red curve) and \ref{fig_setup}\textit{b} (blue), with $N=100$. The distributions are obtained by generating a large number of matrices and separating them in sets having the same number $N_{X}$ of real eigenvalues $\varepsilon=\tan(\phi/2)$, with average spacing $\langle\delta\phi\rangle=2\pi/N_{X}$. A weighted average $P(s)=\sum_{N_{X}}P(s|N_{X})P(N_{X})$ of the spacing distribution within each set is plotted in the figure. The black curves show three analytical spacing distributions.
}
\label{fig_semiPoisson}
\end{figure}

The eigenvalue distribution is known exactly for independent normally distributed matrix elements (the Ginibre ensemble \cite{Gin65,Gir85,Tao10,Bor12}). For large $N$ there are on average $\langle N_{X}\rangle\propto \sqrt{N}$ real eigenvalues \cite{Ede94}. The spacing distribution vanishes as $s^\beta$  for small spacings $s$ (normalized by the average spacing), with $\beta=1$ on the real axis (linear level repulsion) \cite{Leh91,Kan05}. These power laws are derived for uncorrelated matrix elements, but we find numerically \cite{RMT_app} that both the $\sqrt{N}$-scaling (Fig.\ \ref{fig_squareN}) and the linear repulsion (Fig.\ \ref{fig_semiPoisson}) hold for our ensemble of skew-Hamiltonian matrices. 

The linear repulsion for $s\lesssim 1$ crosses over into an exponential tail for $s\gtrsim 1$. As one can see in Fig.\ \ref{fig_semiPoisson}, the semi-Poisson distribution \cite{Shk93,Bog99,Gor01,Gar06} interpolates quite accurately between these small and large-$s$ limits, and describes the numerical RMT results better than either the Poisson distribution of uncorrelated eigenvalues or the Wigner surmise of the Gaussian Orthogonal Ensemble (GOE) \cite{Meh04,For10}. 

The same power laws apply to topologically nontrivial superconductors. We then need reflection matrices $R_{1},R_{2}$ with determinant $-1$, which can be achieved by assuming that a sufficiently large Zeeman energy allows for an unpaired spin channel, and adding this channel as a unit diagonal element to $\Lambda={\rm diag}\,(1,\sigma_{y},\sigma_{y}\ldots\sigma_{y})$. The determinant of the product $R_{1}R_{2}$ remains equal to $+1$, so $O\in{\rm SO}(2N)$ remains special orthogonal, with $N=2M+1$ an odd rather than even integer. Since the eigenvalues of $\cal{M}$ come in complex conjugate pairs, the number $N_{X}$ of distinct real eigenvalues (and hence the number of level crossings) is now also odd rather than even. This even/odd difference does not affect either the $\sqrt{N}$-scaling or the linear repulsion.

So far we considered the geometry of Fig.\ \ref{fig_setup}\textit{b}, with a chaotic mixing of electron and hole degrees of freedom in the quantum dot. In Fig.\ \ref{fig_setup}\textit{a} the quantum dot does not couple electrons and holes, so the random-matrix ensemble is different. The chaotic scattering of electrons in the quantum dot is then described by $N\times N$ reflection and transmission matrices, which together form the unitary scattering matrix $s_{0}$. The scattering matrix for holes, at the Fermi level, is just the complex conjugate $s_{0}^{\ast}$. Instead of the CRE we now have the CUE, the circular unitary ensemble \cite{Dys62}, corresponding to a uniform distribution of $s_{0}\in {\rm U}(2N)$ with the Haar measure of the unitary group. We again find a $\sqrt{N}$ scaling of the number of transitions and a hybrid Wigner-Poisson spacing distribution (red lines in Figs.\ \ref{fig_squareN} and \ref{fig_semiPoisson}) \cite{note3}.

To test these model-independent results of random-matrix theory (RMT), we have performed computer simulations of two microscopic models, one topologically trivial and the other nontrivial. The first model is that of an InSb Josephson junction, similar to that studied in a recent experimental search for Majorana fermions \cite{Rok12}. One crucial difference is that we take a weak perpendicular magnetic field, just a few flux quanta $h/e$ through the junction --- sufficient to break time-reversal symmetry, but not strong enough to induce a transition to a topologically nontrivial state (which would require Zeeman energy comparable to superconducting gap \cite{Ali12}). 

The model Hamiltonian has the Bogoliubov-De Gennes form,
\begin{align}
&{\cal H}=\begin{pmatrix}
H_{0}(\bm{p}-e\bm{A})&\Delta\\
\Delta^{\ast}&-\sigma_{y}H_{0}^{\ast}(-\bm{p}-e\bm{A})\sigma_{y}
\end{pmatrix},\label{HBdGdef}\\
&H_{0}(\bm{p})=\tfrac{1}{2}p^2/m_{\rm eff}+U-E_{\rm F}+\hbar^{-1}\alpha_{\rm so}(\sigma_x p_y-\sigma_y p_x),\nonumber
\end{align}
with electron and hole blocks coupled by the \textit{s}-wave pair potential $\Delta$ at the superconducting contacts. The single-particle Hamiltonian $H_{0}$ contains the Rasba spin-orbit coupling of an InSb quantum well (characteristic length $l_{\rm so}=\hbar^{2}/m_{\rm eff}\alpha_{\rm so}= 0.25\,\mu{\rm m}$) and an electrostatic disorder potential $U$. The vector potential $\bm{A}=(0,Bx,0)$ accounts for the orbital effect of a perpendicular magnetic field $B$ (which we set equal to zero in the superconductors). The Zeeman term has a negligible effect and is omitted. The Fermi energy $E_{\rm F}$ is chosen such that the InSb channel has $N=20$ transverse modes at the Fermi level, including spin. We discretize the model on a two-dimensional square lattice, with disorder potential $U\in(-U_0,U_0)$ chosen randomly and independently on each site. The low-lying energy levels of the resulting tight-binding Hamiltonian are computed \cite{kwant} as a function of the phase difference $\phi$ of the pair potential.

\begin{figure}[tb]
\centerline{\includegraphics[width=0.8\linewidth]{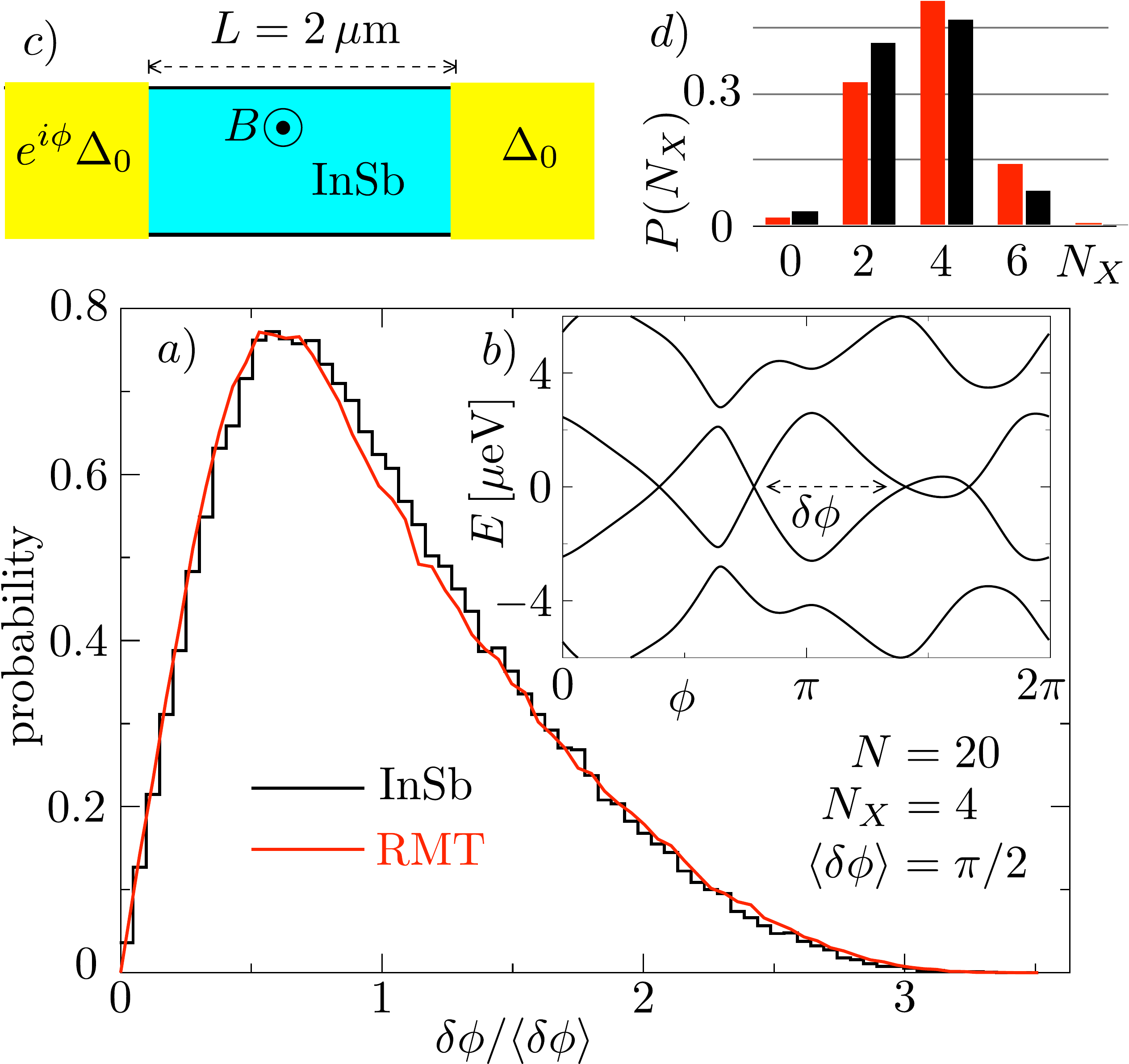}}
\caption{Panel \textit{a}: Spacing distribution of $N_{X}=4$ level crossings in a Josephson junction with $N=20$ transverse modes. The red curve shows the RMT prediction in the geometry of Fig.\ \ref{fig_setup}\textit{a}. The black histogram is a model calculation \cite{InSbparam} for a disordered InSb channel in a perpendicular magnetic field, sampled over different impurity configurations (the inset \textit{b} shows level crossings for one sample and panel \textit{c} shows the geometry). Panel \textit{d} compares the probability of $N_{X}$ level crossings for $N=20$ in the RMT calculation (red curve) and in the InSb model (black). 
}
\label{fig_wire}
\end{figure}

In Fig.\ \ref{fig_wire} we compare the results of the InSb model calculation \cite{InSbparam} with the RMT predictions in the quantum-dot geometry of Fig.\ \ref{fig_setup}\textit{a}. The disordered InSb channel lacks the point contact coupling of a quantum dot, so the scattering is not fully chaotic and no precise agreement with the RMT calculations is to be expected. Indeed, the probabilities $P(N_{X})$ to have $N_{X}$ level crossings for $N=20$ modes, shown in Fig.\ \ref{fig_wire}\textit{d}, agree only qualitatively. Still, the spacing distributions, shown in Fig.\ \ref{fig_wire}\textit{a} for $N_{X}=4$, are in remarkable agreement --- \textit{without any adjustable parameter}.

The second microscopic model that we have studied is topologically nontrivial: the quantum spin-Hall (QSH) insulator in a InAs/GaSb quantum well \cite{Liu08,Kne12}. The Hamiltonian still has the Bogoliubov-De Gennes form \eqref{HBdGdef}, but now $H_{0}$ is the four-band Bernevig-Hughes-Zhang Hamiltonian \cite{Ber06}. The quantum dot is formed using the method of Ref.\ \onlinecite{Mi13}, by locally pushing the conduction band below the Fermi level by means of a gate electrode. The QSH insulator has a single conducting mode at the edge \cite{Has10,Qi11}, so $N=1$ and our large-$N$ RMT is not directly applicable. Still, as shown in Fig.\ \ref{fig_dot}, a linear repulsion at small spacings still applies if we count the level crossings as a function of the chemical potential in the quantum dot --- demonstrating the universality of this effect.

In conclusion, we have discovered a statistical correlation in the fermion-parity switches of a Josephson junction. The spacing distribution of these topological phase transitions has a universal form, a hybrid of the Wigner and Poisson distributions, decaying linearly at small spacings and exponentially at large spacings. Such a hybrid (semi-Poisson or ``mermaid'') distribution is known from Anderson phase transitions \cite{Shk93,Bog99}, where it signals a fractal structure of wave functions. It would be interesting for further theoretical work to investigate whether this self-similar structure appears here as well. Experimentally, it would be of interest to search for the repulsion of level crossings by tunnel spectroscopy \cite{Cha12}.

We thank A. R. Akhmerov for discussions and help with the random-matrix calculations. This work was supported by the Dutch Science Foundation NWO/FOM, by an ERC Advanced Investigator Grant, by the EU network NanoCTM, and by the China Scholarship Council.
\newpage

\appendix
\begin{widetext}
\section{Eigenvalue statistics of real non-Hermitian matrices}
\label{RMTdetails}

In this Appendix we collect numerical results for the statistics of the eigenvalues of a real non-Hermitian matrix. We compare five different ensembles, summarized in Table \ref{tab:overview}, to show the universality of the square-root law for the average number of real eigenvalues and for the hybrid Wigner-Poisson spacing distribution on the real axis.

\begin{table}[h]
  \centering
  \begin{tabular}{ c || c | c | c | c | c}
    ensemble & symmetry & measure & matrix size & distinct eigenv. & $\lim_{N\rightarrow\infty}N_X^2/N$   \\
    \hline                                                                                                                                                              
 \textit{skew-CRE} & ${\cal M}^{\rm T}=-J{\cal M}J$ & Haar on ${\rm SO}(2N)$ & $2N\times 2N$ & $N$ & $\approx 4/\pi$ \\
 \textit{skew-CUE} & ${\cal M}^{\rm T}=-J{\cal M}J$ & Haar on ${\rm U}(2N)$ & $2N\times 2N$ & $N$ & $\approx 2/\pi$ \\
 \textit{skew-GOE} & ${\cal M}^{\rm T}=-J{\cal M}J$ & Gaussian &$2N\times 2N$ & $N$ & $\approx \pi/6$ \\
 \textit{Hamiltonian} & ${\cal M}^{\rm T}=J{\cal M}J$ & Gaussian & $2N\times 2N$ & $2N$ & $\approx 4/\pi$ \\
 \textit{Ginibre} & --- & Gaussian & $N\times N$ & $N$ & $= 2/\pi$
\end{tabular}
  \caption{Overview of the five ensembles of real non-Hermitian matrices considered here. The last column summarizes the $\sqrt{N}$ law for the number $N_X$ of distinct real eigenvalues. The coefficients are conjectures based on numerical data, except for the Ginibre ensemble, where it is an exact result from Ref.\ \onlinecite{Ede94}.
}
\label{tab:overview}
\end{table}
\end{widetext}

\subsection{Skew-Hamiltonian ensembles}
\label{skewHensemble}

The definition ${\cal M}^{\rm T}=-J{\cal M}J$ of a skew-Hamiltonian real matrix implies that it can be written in the form ${\cal M}=AJ$, with $J$ the fundamental antisymmetric (= skew-symmetric) matrix
\begin{equation}
J=\begin{pmatrix}
\emptyset_{N}&\openone_{N}\\
-\openone_{N}&\emptyset_{N}
\end{pmatrix},\label{Jdef}
\end{equation}
and $A=-A^{\rm T}$ real antisymmetric. The matrices $J$ and $A$ have dimension $2N\times 2N$. The subblocks $\emptyset_{N}$ and $\openone_{N}$ are $N\times N$ diagonal matrices with, respectively, $0$ and $1$ on the diagonal. 

Barring accidental degeneracies, the $2N\times 2N$ matrix ${\cal M}$ has $N$ distinct eigenvalues, each with multiplicity two, symmetrically arranged around the real axis. We seek the probability $P(N_{X})$ that there are $N_{X}$ distinct real eigenvalues. This probability is only nonzero if $N_{X}=1,3,5,\ldots N$ for $N$ odd, or $N_{X}=0,2,4,\ldots N$ for $N=2M$ even. As worked out in the main text, the real eigenvalues $\varepsilon_{n}$ identify the phases $\phi_{n}=2\,{\rm arctan}\,\varepsilon_{n}$ of a level crossing in the quantum-dot Josephson junction. 

Tables \ref{tab:CRE},\ref{tab:CUE}, and \ref{tab:GOE} list numerical results \cite{RMT_app} for  the following three ensembles of skew-Hamiltonian matrices:\medskip

$\bullet$ \textit{skew-CRE:} the skew-Hamiltonian ensemble derived from the circular real ensemble (CRE).\smallskip\\
This ensemble applies to the geometry of Fig.\ \ref{fig_setup}\textit{b}. Starting from a matrix $O$ that is uniformly distributed with the Haar measure in ${\rm SO}(2N)$, we construct the skew-Hamiltonian matrix
\begin{equation}
{\cal M}=(1-O)(1+O)^{-1}J.\label{MOJ}
\end{equation}

$\bullet$ \textit{skew-CUE:} the skew-Hamiltonian ensemble derived from the circular unitary ensemble (CUE).\smallskip\\
This ensemble applies to the geometry of Fig.\ \ref{fig_setup}\textit{a}. We start from a scattering matrix $s_{0}$ of the quantum dot that is uniformly distributed with the Haar measure in ${\rm U}(2N)$. The matrix $s_{0}$ has the block structure
\begin{equation}
s_{0}=\begin{pmatrix}
r'&t\\
t'&r
\end{pmatrix},\label{Urt}
\end{equation}
with $N\times N$ transmission and reflection matrices $t,t',r,r'$. 

Since the Haar measure is the same in any basis, we are free to choose a basis for $s_{0}$ such that the Andreev reflection matrix at $\phi=0$ is the unit matrix. The electron-hole reflection matrices $R_{1}$ and $R_{2}$ are then given by
\begin{align}
&R_{1}=\begin{pmatrix}
\emptyset_{N}&\openone_{N}\\
\openone_{N}&\emptyset_{N}
\end{pmatrix},\;\;R_{2}=\begin{pmatrix}
r_{ee}&r_{he}^{\ast}\\
r_{he}&r_{ee}^{\ast}
\end{pmatrix},\label{R1R2def}\\
&r_{ee}=r'+tr^{\ast}(1-rr^{\ast})^{-1}t',\;\;r_{he}=t^{\ast}(1-rr^{\ast})^{-1}t'.
\label{R2s0def}
\end{align}

We then construct the matrix $O\in{\rm SO}(2N)$ from the matrix product
\begin{equation}
O=\Omega^{\dagger}R_{2}R_{1}\Omega,\;\;\Omega=\frac{1}{\sqrt{2}}\begin{pmatrix}
\openone_{N}&i\,\openone_{N}\\
\openone_{N}&-i\,\openone_{N}
\end{pmatrix},\label{OR2R1U}
\end{equation}
and from $O$ we arrive at the skew-Hamiltonian matrix ${\cal M}$ via Eq.\ \eqref{MOJ}.\medskip

$\bullet$ \textit{skew-GOE:} the skew-Hamiltonian ensemble derived from the Gaussian orthogonal ensemble (GOE).\smallskip\\
This ensemble does not correspond to a scattering problem and is included for comparison. The skew-Hamiltonian matrix ${\cal M}=AJ$ is constructed by taking independent Gaussian distributions for the upper diagonal elements of the $2N\times 2N$ antisymmetric matrix $A$,
\begin{equation}
P(A)\propto\left(-\tfrac{1}{2}\sum_{n<m}A_{nm}^{2}\right).\label{PA}
\end{equation}

\subsection{Hamiltonian and Ginibre ensembles}
\label{Hamiltonian}

In the skew-Hamiltonian ensembles all eigenvalues are two-fold degenerate, which in the context of the Josephson junction signifies that a level crossing is a degeneracy point for a pair of Andreev levels. We would like to see to what extent this special feature plays a role in the statistics, so we compare with two ensembles where all eigenvalues are distinct --- but which still show an accumulation of eigenvalues on the real axis.\medskip

$\bullet$ \textit{Hamiltonian ensemble}\smallskip\\
A $2N\times 2N$ real matrix ${\cal M}$ is called Hamiltonian if it satisfies ${\cal M}^{\rm T}=J{\cal M}J$, which means that it can be written in the form ${\cal M}=HJ$ with $H=H^{\rm T}$ real symmetric. We draw $H$ from the Gaussian orthogonal ensemble,
\begin{equation}
P(H)\propto\exp\left(-\tfrac{1}{2}\sum_{n}H_{nn}^{2}-\sum_{n<m}H_{nm}^{2}\right),\label{PGOE} 
\end{equation}
to produce an ensemble of random Hamiltonian matrices\footnote{This ensemble was suggested as a research topic by Austen Lamacraft at {\tt mathoverflow.net} (question 120397).} 
${\cal M}=HJ$.

The $2N$ eigenvalues are all distinct, barring accidental degeneracies. They are symmetrically arranged around the real and imaginary axis, and they accumulate on both these axes. (This is easily understood by noting that the square of a Hamiltonian matrix is skew-Hamiltonian.) The probability $P(N_{X})$ that there are $N_{X}$ real eigenvalues is only nonzero if $N_{X}=0,2,4,\ldots 2N$, irrespective of whether $N$ is even or odd, see Table \ref{tab:Hamiltonian}\textit{a}. The same appplies to the probability $P(N_{Y})$ that there are $N_{Y}$ imaginary eigenvalues, listed in Table \ref{tab:Hamiltonian}\textit{b}.\medskip

$\bullet$ \textit{Ginibre ensemble}\smallskip\\
The four ensembles considered so far are only defined for even dimensional matrices (size $2N\times 2N$). The Ginibre ensemble of real matrices \cite{Gin65} is defined for both even and odd dimensions, so we denote its size by $N\times N$. The $N^2$ matrix elements are drawn independently from the same Gaussian,
\begin{equation}
P(H)\propto\exp\left(-\tfrac{1}{2}\sum_{n,m}H_{nm}^{2}\right).\label{PGinibre} 
\end{equation}
The $N$ eigenvalues are all distinct, symmetrically arranged around the real axis, with accumulation only on that axis. The probability $P(N_{X})$ is only nonzero if $N_{X}=1,3,5,\ldots N$ for $N$ odd, or $N_{X}=0,2,4,\ldots N$ for $N=2M$ even. 

The Ginibre ensemble is the only ensemble of real non-Hermitian matrices where the probability of real eigenvalues is known analytically \cite{Ede94}, listed in Table \ref{tab:Ginibre}.

\begin{table}[tb]
  \centering
  \begin{tabular}{ c | c || c | c | c | c | c | c | c | c | c  }
  \multicolumn{11}{c}{\textit{skew-CRE}}                                                        \\
  \hline
        &                        & \multicolumn{9}{|c}{$P(N_{X})$ for $N_{X}$ equal to:}        \\
    $N$ & $\langle N_{X}\rangle$ & $0$  & $1$  & $2$  & $3$  & $4$  & $5$  & $6$  & $7$  & $8$  \\
    \hline
    1   & 1                      & 0    & 1    & 0    & 0    & 0    & 0    & 0    & 0    & 0    \\
    2   & 1.50                   & 0.25 & 0    & 0.75 & 0    & 0    & 0    & 0    & 0    & 0    \\   
    3   & 1.88                   & 0    & 0.56 & 0    & 0.44 & 0    & 0    & 0    & 0    & 0    \\  
    4   & 2.19                   & 0.11 & 0    & 0.70 & 0    & 0.20 & 0    & 0    & 0    & 0    \\  
    5   & 2.46                   & 0    & 0.34 & 0    & 0.59 & 0    & 0.07 & 0    & 0    & 0    \\  
    6   & 2.70                   & 0.05 & 0    & 0.56 & 0    & 0.37 & 0    & 0.02 & 0    & 0    \\  
    7   & 2.93                   & 0    & 0.22 & 0    & 0.60 & 0    & 0.17 & 0    & 0.00 & 0    \\  
    8   & 3.15                   & 0.03 & 0    & 0.43 & 0    & 0.47 & 0    & 0.06 & 0    & 0.00 \\  
    9   & 3.34                   & 0    & 0.15 & 0    & 0.56 & 0    & 0.28 & 0    & 0.02 & 0    \\  
    10  & 3.52                   & 0.02 & 0    & 0.34 & 0    & 0.52 & 0    & 0.13 & 0    & 0.00 \\
  \end{tabular}
  \caption{Probability $P(N_{X})$ that the $N$-mode quantum-dot Josephson junction in the geometry of Fig.\ \ref{fig_setup}\textit{b} has $N_{X}$ level crossings in a $2\pi$ phase interval. The data is calculated by generating a large number of random reflection matrices $O$, uniformly distributed in ${\rm SO}(2N)$, and identifying $N_{X}$ with the number of distinct real eigenvalues of the skew-Hamiltonian matrix ${\cal M}=(1-O)(1+O)^{-1}J$. The average $\langle N_{X}\rangle$ is listed in the second column.}
  \label{tab:CRE}
\end{table}

\begin{table}[tb]
  \centering
  \begin{tabular}{ c | c || c | c | c | c | c | c | c | c | c}
    \multicolumn{11}{c}{\textit{skew-CUE}}                                                      \\
  \hline
        &                        & \multicolumn{9}{|c}{$P(N_{X})$ for $N_{X}$ equal to:}        \\
    $N$ & $\langle N_{X}\rangle$ & $0$  & $1$  & $2$  & $3$  & $4$  & $5$  & $6$  & $7$  & $8$  \\
    \hline
    1   & 1                      & 0    & 1    & 0    & 0    & 0    & 0    & 0    & 0    & 0    \\
    2   & 1.20                   & 0.40 & 0    & 0.60 & 0    & 0    & 0    & 0    & 0    & 0    \\   
    3   & 1.43                   & 0    & 0.78 & 0    & 0.22 & 0    & 0    & 0    & 0    & 0    \\  
    4   & 1.63                   & 0.23 & 0    & 0.73 & 0    & 0.04 & 0    & 0    & 0    & 0    \\  
    5   & 1.82                   & 0    & 0.60 & 0    & 0.40 & 0    & 0.01 & 0    & 0    & 0    \\  
    6   & 1.98                   & 0.15 & 0    & 0.72 & 0    & 0.13 & 0    & 0.00 & 0    & 0    \\  
    7   & 2.14                   & 0    & 0.46 & 0    & 0.51 & 0    & 0.03 & 0    & 0.00 & 0    \\  
    8   & 2.28                   & 0.10 & 0    & 0.67 & 0    & 0.23 & 0    & 0.00 & 0    & 0.00 \\  
    9   & 2.42                   & 0    & 0.36 & 0    & 0.57 & 0    & 0.07 & 0    & 0.00 & 0    \\  
    10  & 2.54                   & 0.07 & 0    & 0.60 & 0    & 0.32 & 0    & 0.01 & 0    & 0.00 \\
  \end{tabular}
  \caption{Same as Table \ref{tab:CRE}, but for the geometry of Fig.\ \ref{fig_setup}\textit{a}. The random-matrix ensemble now corresponds to the uniform distribution of $s_{0}\in {\rm U}(2N)$.}
  \label{tab:CUE}
\end{table}

\begin{table}[tb]
  \centering
  \begin{tabular}{ c | c || c | c | c | c | c | c | c | c | c  }
    \multicolumn{11}{c}{\textit{skew-GOE}}                                                      \\
  \hline
        &                        & \multicolumn{9}{|c}{$P(N_{X})$ for $N_{X}$ equal to:}        \\
    $N$ & $\langle N_{X}\rangle$ & $0$  & $1$  & $2$  & $3$  & $4$  & $5$  & $6$  & $7$  & $8$  \\
    \hline                                                                                                                                                       
    1   & 1                      & 0    & 1    & 0    & 0    & 0    & 0    & 0    & 0    & 0    \\
     2  & 1.29                   & 0.35 & 0    & 0.65 & 0    & 0    & 0    & 0    & 0    & 0    \\
     3  & 1.52                   & 0    & 0.74 & 0    & 0.26 & 0    & 0    & 0    & 0    & 0    \\
     4  & 1.71                   & 0.21 & 0    & 0.73 & 0    & 0.06 & 0    & 0    & 0    & 0    \\
     5  & 1.88                   & 0    & 0.57 & 0    & 0.42 & 0    & 0.01 & 0    & 0    & 0    \\
     6  & 2.03                   & 0.14 & 0    & 0.71 & 0    & 0.15 & 0    & 0.00 & 0    & 0    \\
     7  & 2.16                   & 0    & 0.45 & 0    & 0.51 & 0    & 0.03 & 0    & 0.00 & 0    \\
     8  & 2.30                   & 0.10 & 0    & 0.66 & 0    & 0.24 & 0    & 0.00 & 0    & 0.00 \\
     9  & 2.42                   & 0    & 0.36 & 0    & 0.57 & 0    & 0.07 & 0    & 0.00 & 0    \\
     10 & 2.54                   & 0.07 & 0    & 0.60 & 0    & 0.31 & 0    & 0.01 & 0    & 0.00 \\
\end{tabular}
  \caption{Probability of $N_{X}$ distinct real eigenvalues of the skew-Hamiltonian matrix $AJ$, in the ensemble of $2N\times 2N$ antisymmetric matrices $A$ with independent real Gaussian elements on the upper diagonal.}
  \label{tab:GOE}
\end{table}

\begin{table*}[tb]
  \centering
  \begin{tabular}{ c | c || c | c | c | c | c | c | c | c | c|c|c  }
    \multicolumn{13}{c}{\textit{a) Hamiltonian ensemble (real eigenvalues)}}                                                   \\
  \hline
        &                        & \multicolumn{11}{|c}{$P(N_{X})$ for $N_{X}$ equal to:}                \\
    $N$ & $\langle N_{X}\rangle$ & $0$  & $1$ & $2$  & $3$ & $4$  & $5$ & $6$  & $7$ & $8$  & $9$ & $10$ \\
    \hline                                                                                                                                                       
   1    & 1.42                   & 0.29 & 0   & 0.71 & 0   & 0    & 0   & 0    & 0   & 0    & 0   & 0    \\
     2  & 1.71                   & 0.40 & 0   & 0.36 & 0   & 0.25 & 0   & 0    & 0   & 0    & 0   & 0    \\
     3  & 2.11                   & 0.21 & 0   & 0.57 & 0   & 0.18 & 0   & 0.04 & 0   & 0    & 0   & 0    \\
     4  & 2.41                   & 0.24 & 0   & 0.35 & 0   & 0.35 & 0   & 0.04 & 0   & 0.00 & 0   & 0    \\
     5  & 2.67                   & 0.15 & 0   & 0.48 & 0   & 0.25 & 0   & 0.11 & 0   & 0.01 & 0   & 0.00 \\
     6  & 2.91                   & 0.17 & 0   & 0.33 & 0   & 0.39 & 0   & 0.09 & 0   & 0.02 & 0   & 0.00 \\
     7  & 3.14                   & 0.11 & 0   & 0.42 & 0   & 0.28 & 0   & 0.16 & 0   & 0.02 & 0   & 0.00 \\
     8  & 3.35                   & 0.13 & 0   & 0.29 & 0   & 0.40 & 0   & 0.14 & 0   & 0.04 & 0   & 0.00 \\
     9  & 3.55                   & 0.09 & 0   & 0.35 & 0   & 0.30 & 0   & 0.22 & 0   & 0.04 & 0   & 0.00 \\
     10 & 3.73                   & 0.10 & 0   & 0.26 & 0   & 0.40 & 0   & 0.18 & 0   & 0.06 & 0   & 0.00 \\
\end{tabular}\quad\quad
 \begin{tabular}{ c | c || c | c | c | c | c | c | c | c | c|c|c  }
     \multicolumn{13}{c}{\textit{b) Hamiltonian ensemble (imaginary eigenvalues)}}                                                   \\
  \hline
        &                        & \multicolumn{11}{|c}{$P(N_{Y})$ for $N_{Y}$ equal to:}                \\
    $N$ & $\langle N_{Y}\rangle$ & $0$  & $1$ & $2$  & $3$ & $4$  & $5$ & $6$  & $7$ & $8$  & $9$ & $10$ \\
    \hline                                                                                                                                                                       
    1   & 0.58                                & 0.71     & 0      & 0.29     & 0      & 0      & 0      & 0      & 0      & 0      & 0      & 0       \\
     2  & 0.88                                & 0.6      & 0      & 0.36     & 0      & 0.04     & 0      & 0      & 0      & 0      & 0      & 0       \\
     3  & 1.11                                & 0.53     & 0      & 0.38     & 0      & 0.08     & 0      & 0.00      & 0      & 0      & 0      & 0       \\
     4  & 1.29                                & 0.49     & 0      & 0.39     & 0      & 0.11     & 0      & 0.01     & 0      & 0.00      & 0      & 0       \\
     5  & 1.47                                & 0.45     & 0      & 0.39     & 0      & 0.15     & 0      & 0.02     & 0      & 0.00      & 0      & 0.00       \\
     6  & 1.61                                & 0.41     & 0      & 0.4      & 0      & 0.17     & 0      & 0.02     & 0      & 0.00      & 0      & 0.00       \\
     7  & 1.74                                & 0.39     & 0      & 0.39     & 0      & 0.19     & 0      & 0.03     & 0      & 0.00      & 0      & 0.00       \\
     8  & 1.88                                & 0.36     & 0      & 0.39     & 0      & 0.2      & 0      & 0.04     & 0      & 0.00      & 0      & 0.00       \\
     9  & 1.99                                & 0.34     & 0      & 0.38     & 0      & 0.22     & 0      & 0.05     & 0      & 0.01     & 0      & 0.00       \\
     10 & 2.11                                & 0.32     & 0      & 0.38     & 0      & 0.23     & 0      & 0.06     & 0      & 0.01     & 0      & 0.00       \\
\end{tabular}
  \caption{Probability $P(N_{X})$ of $N_{X}$ real eigenvalues, and probability $P(N_{Y})$ of $N_{Y}$ imaginary eigenvalues of the Hamiltonian matrix $HJ$, constructed from the $2N\times 2N$ symmetric matrix $H$ in the GOE.}
  \label{tab:Hamiltonian}
\end{table*}

\begin{table}[tb]
  \centering
  \begin{tabular}{ c | c || c | c | c | c | c | c | c | c | c }
      \multicolumn{11}{c}{\textit{Ginibre ensemble}}                                            \\
  \hline
        &                        & \multicolumn{9}{|c}{$P(N_{X})$ for $N_{X}$ equal to:}        \\
    $N$ & $\langle N_{X}\rangle$ & $0$  & $1$  & $2$  & $3$  & $4$  & $5$  & $6$  & $7$  & $8$  \\
    \hline                                                                                                                                                              
    1   & 1                      & 0    & 1    & 0    & 0    & 0    & 0    & 0    & 0    & 0    \\
     2  & 1.41                   & 0.29 & 0    & 0.71 & 0    & 0    & 0    & 0    & 0    & 0    \\
     3  & 1.71                   & 0    & 0.65 & 0    & 0.35 & 0    & 0    & 0    & 0    & 0    \\
     4  & 1.94                   & 0.15 & 0    & 0.72 & 0    & 0.12 & 0    & 0    & 0    & 0    \\
     5  & 2.15                   & 0    & 0.46 & 0    & 0.51 & 0    & 0.03 & 0    & 0    & 0    \\
     6  & 2.33                   & 0.10 & 0    & 0.65 & 0    & 0.25 & 0    & 0.01 & 0    & 0    \\
     7  & 2.50                   & 0    & 0.34 & 0    & 0.58 & 0    & 0.08 & 0    & 0.00 & 0    \\
     8  & 2.66                   & 0.06 & 0    & 0.57 & 0    & 0.35 & 0    & 0.02 & 0    & 0.00 \\
     9  & 2.79                   & 0    & 0.26 & 0    & 0.59 & 0    & 0.15 & 0    & 0.00 & 0    \\
     10 & 2.92                   & 0.04 & 0    & 0.49 & 0    & 0.42 & 0    & 0.04 & 0    & 0.00 \\
\end{tabular}
  \caption{Probability of $N_{X}$ real eigenvalues of an $N\times N$ matrix with all matrix elements draw independently from the same  Gaussian distribution. These are analytical results from Ref.\ \cite{Ede94}. We have checked that our numerics gives the same numbers, to two decimal places.}
  \label{tab:Ginibre}
\end{table}

\subsection{Circular law}
\label{accumulation}

\begin{figure*}[tb]
\centerline{\includegraphics[width=0.8\linewidth]{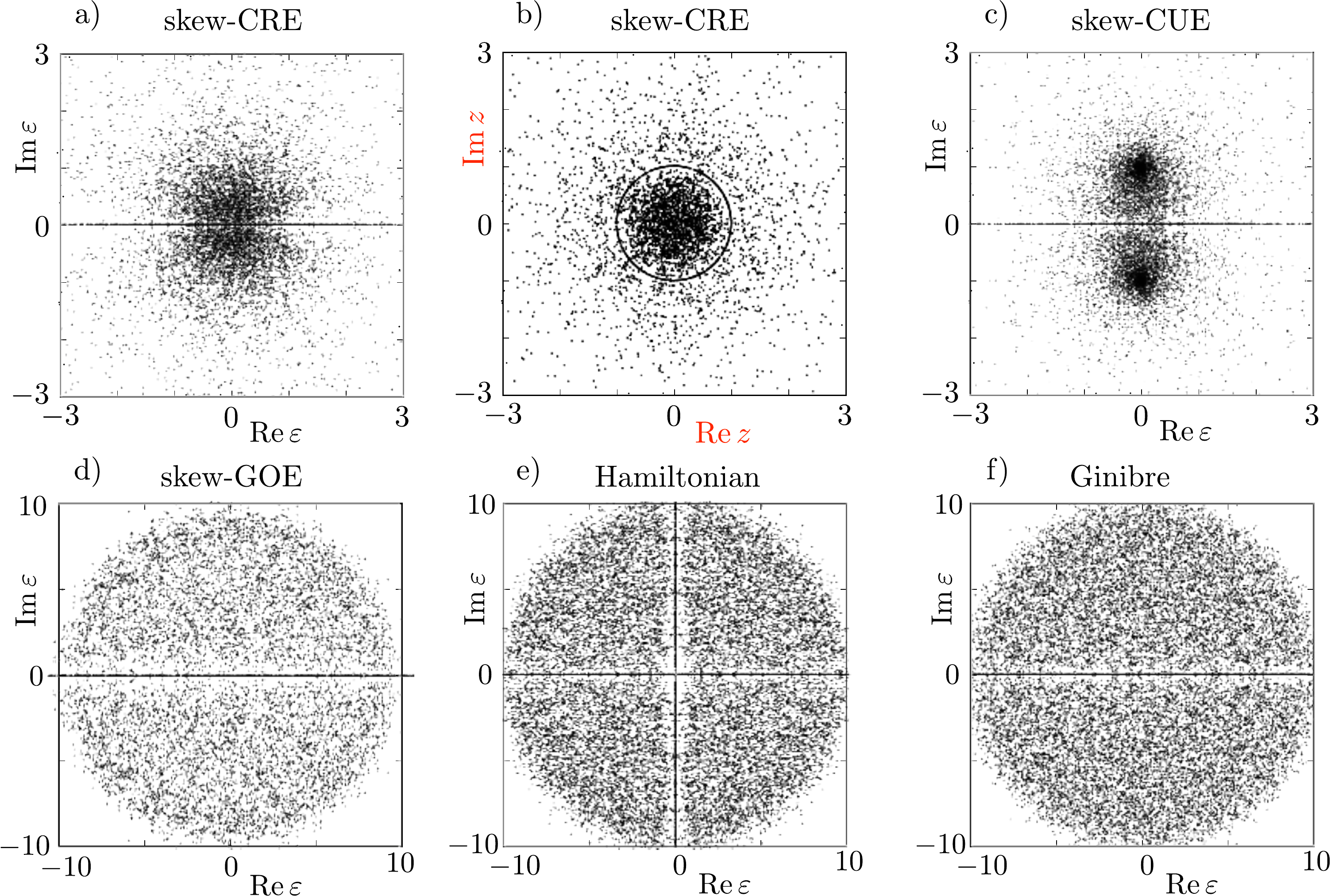}}
\caption{Eigenvalues of $200$ non-Hermitian real matrices of size $100\times 100$, for the different ensembles. All panels show the complex eigenvalues $\varepsilon$ themselves, except panel \textit{b)}, where we have made the conformal transformation $z=(1+i\varepsilon)/(1-i\varepsilon)$ that maps the real line onto the unit circle. 
}
\label{fig_circleLaw}
\end{figure*}

Fig.\ \ref{fig_circleLaw} shows a scatter plot of the eigenvalues in these random-matrix ensembles. The accumulation of eigenvalues on the real axis is clearly visible. (In the Hamiltonian ensemble the eigenvalues accumulate also on the imaginary axis.)

For the three ensembles of $N'\times N'$ matrices with independent Gaussian matrix elements --- the skew-GOE ($N'=2N$), Hamiltonian ($N'=2N$), and Ginibre ensemble ($N'=N$) --- we find that the complex eigenvalues $\varepsilon_n$ are approximately uniformly distributed within a circle of radius $\sqrt{N'}$ in the complex $\varepsilon$-plane. This is the celebrated \textit{circular law} \cite{Gir85}, proven \cite{Leh91,Tao10,Bor12} for the Ginibre ensemble in the large $N'$-limit:
\begin{equation}
\lim_{N'\rightarrow\infty}\frac{\pi}{N'}\left\langle\sum_{n=1}^{N'}\delta\left(z-\varepsilon_{n}/\sqrt{N'}\right)\right\rangle=\begin{cases}
1&{\rm if}\;\;|z|<1,\\
0&{\rm if}\;\;|z|>1.
\end{cases}
\label{circularlaw}
\end{equation}
Our numerics suggests that the same circular law applies to the skew-GOE and Hamiltonian ensembles. Because each eigenvalue is twofold degenerate in the skew-GOE, they appear less dense in the scatter plot --- compare Figs.\ \ref{fig_circleLaw}\textit{d} and \ref{fig_circleLaw}\textit{f}.

\begin{figure*}[tb]
\centerline{\includegraphics[width=0.7\linewidth]{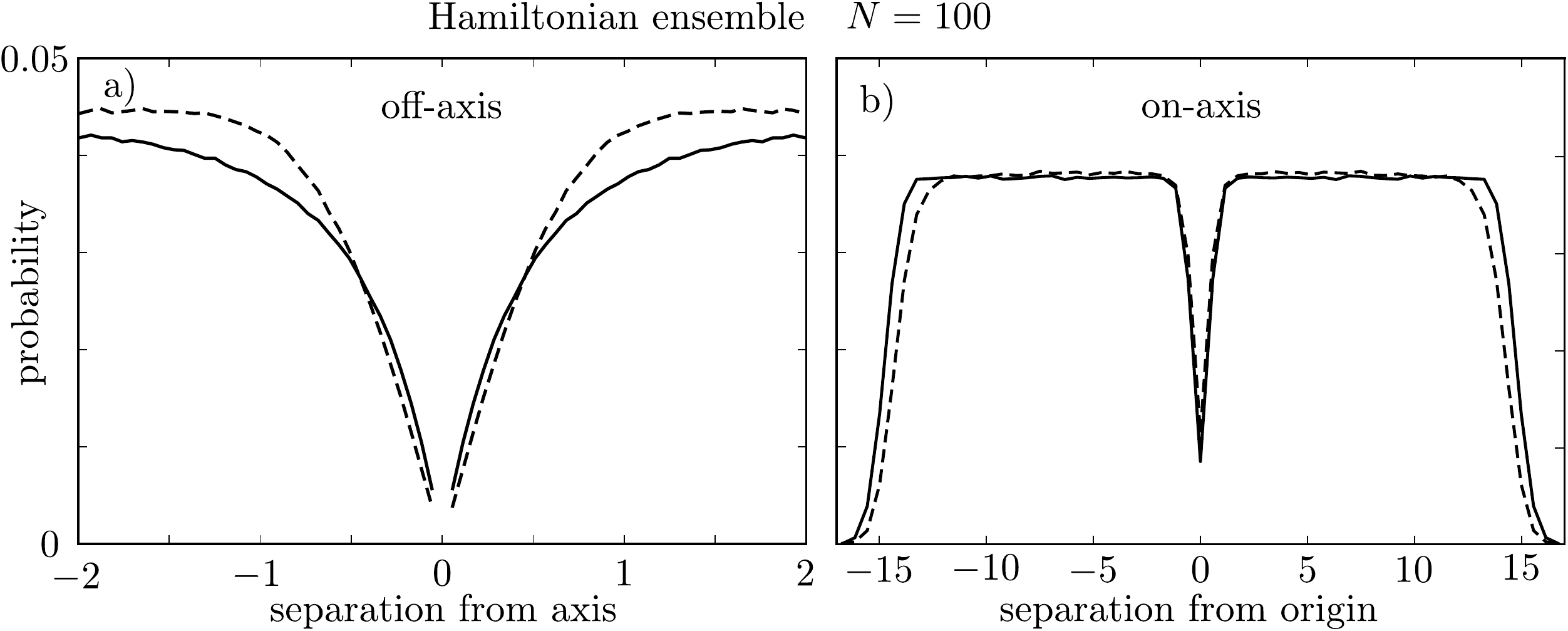}}
\caption{Eigenvalue probability densities for the ensemble of Hamiltonian matrices of size $200\times 200$: \textit{a)} In the depletion zone near the real axis (solid curve) and near the imaginary axis (dashed), as a function of the separation from the axis; \textit{b)} On the real axis (solid curve) and on the imaginary axis (dashed), as a function of the separation from the origin. Each probability density is normalized to unity, when the separation is integrated from $-\infty$ to $\infty$.
}
\label{fig_depletion}
\end{figure*}

The narrow depletion zones surrounding the real axis in Figs.\ \ref{fig_circleLaw}\textit{d,e,f} (and also surrounding the imaginary axis in Fig.\ \ref{fig_circleLaw}\textit{e}) are a finite-$N'$ correction to the circular law, corresponding to a linearly vanishing eigenvalue density --- see Fig.\ \ref{fig_depletion}\textit{a}. The eigenvalue density $\rho(\varepsilon)$ on the real axis is approximately uniform for $-\sqrt{N'}<\varepsilon<\sqrt{N'}$. In the Hamiltonian ensemble also the density on the imaginary axis is approximately uniform in the same interval, except within a distance of order unity from the origin, where the $\pm\varepsilon$ symmetry produces a linear level repulsion --- see Fig.\ \ref{fig_depletion}\textit{b}.

The circular law evidently does not apply to the two ensembles of skew-Hamiltonian matrices constructed from the Haar measure for unitary or orthogonal matrices, see Figs.\ \ref{fig_circleLaw}\textit{a,c}. The eigenvalue density in these ensembles lacks rotational symmetry, which can be restored by the conformal transformation
\begin{equation}
\varepsilon\mapsto z=\frac{1+i\varepsilon}{1-i\varepsilon},\label{epsilonzmapping}
\end{equation}
see Fig.\ \ref{fig_circleLaw}\textit{b}. This transformation is the analytic continuation of $\varepsilon=\tan(\phi/2)$, $z=e^{i\phi}$, to complex $\phi$. The real axis in the complex $\varepsilon$-plane is mapped onto the unit circle in the complex $z$-plane. The rotational symmetry on the unit circle implies that the real $\varepsilon$'s have a Lorentzian density profile,
\begin{equation}
\rho(\varepsilon)=\frac{1}{2\pi}\frac{d\phi}{d\varepsilon}=\frac{1}{\pi}\,\frac{1}{1+\varepsilon^{2}},\;\;\varepsilon\in\mathbb{R}.\label{rhoepsilon}
\end{equation}

\subsection{Square-root law}
\label{squareroot}

\begin{figure}[tb]
\centerline{\includegraphics[width=0.9\linewidth]{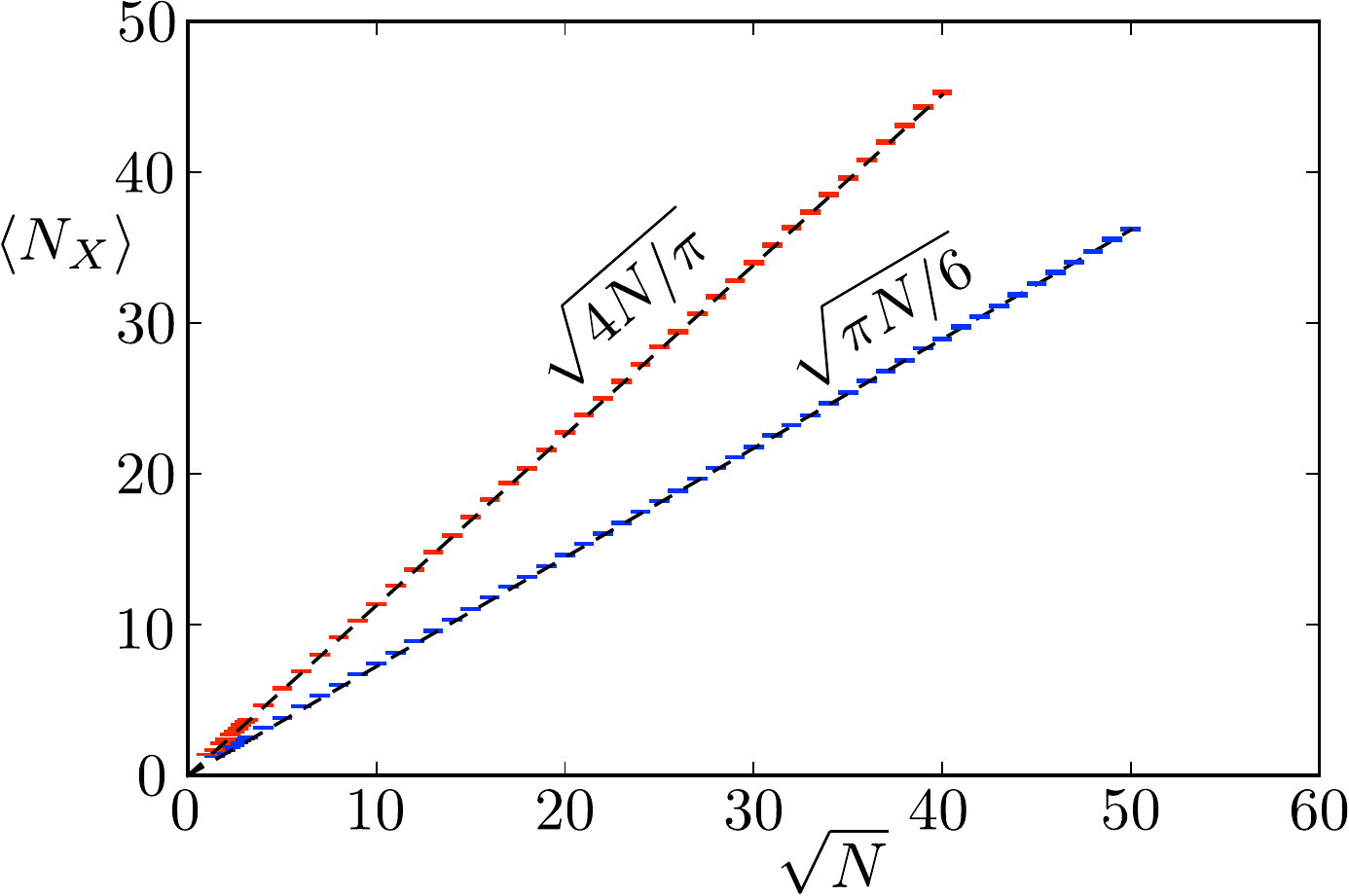}}
\caption{Average number $\langle N_{X}\rangle$ of distinct real eigenvalues, for $2N\times 2N$ random matrices with Gaussian distributed elements and Hamiltonian symmetry (red data points) or skew-Hamiltonian symmetry (blue data points). The formulas given by the dashed lines describe the large-$N$ data well, but they are not derived analytically.
}
\label{fig_squareN_app}
\end{figure}

The square-root law says that the average number of real eigenvalues of a large real random matrix scales as the square root of the size of the matrix,
\begin{equation}
\langle N_{X}\rangle=\sqrt{c_{1}N}+c_{0}+{\cal O}(N^{-1/2}).\label{squarerootC1C0}
\end{equation}
This scaling has been derived for the Ginibre ensemble of $N\times N$ real Gaussian matrices \cite{Ede94}, where $c_{1}=2/\pi$ and $c_{0}=1/2$. All the ensembles considered here follow the same $\sqrt{N}$ scaling, with different coefficients: see Fig.\ \ref{fig_squareN}, for the skew-CUE and skew-CRE, and Fig.\ \ref{fig_squareN_app}, for the skew-GOE and Hamiltonian ensembles.

In the ensemble of $2N\times 2N$ Hamiltonian matrices we denote by $N_{Y}$ the number of purely imaginary eigenvalues. The average $\langle N_{Y}\rangle\approx 0.717\,\sqrt{N}$ scales with the same power of $N$ but a smaller slope than $\langle N_{X}\rangle\approx 1.128\,\sqrt{N}$.

\subsection{Spacing distribution of real eigenvalues}
\label{linearrep}

\begin{figure}[tb]
\centerline{\includegraphics[width=0.9\linewidth]{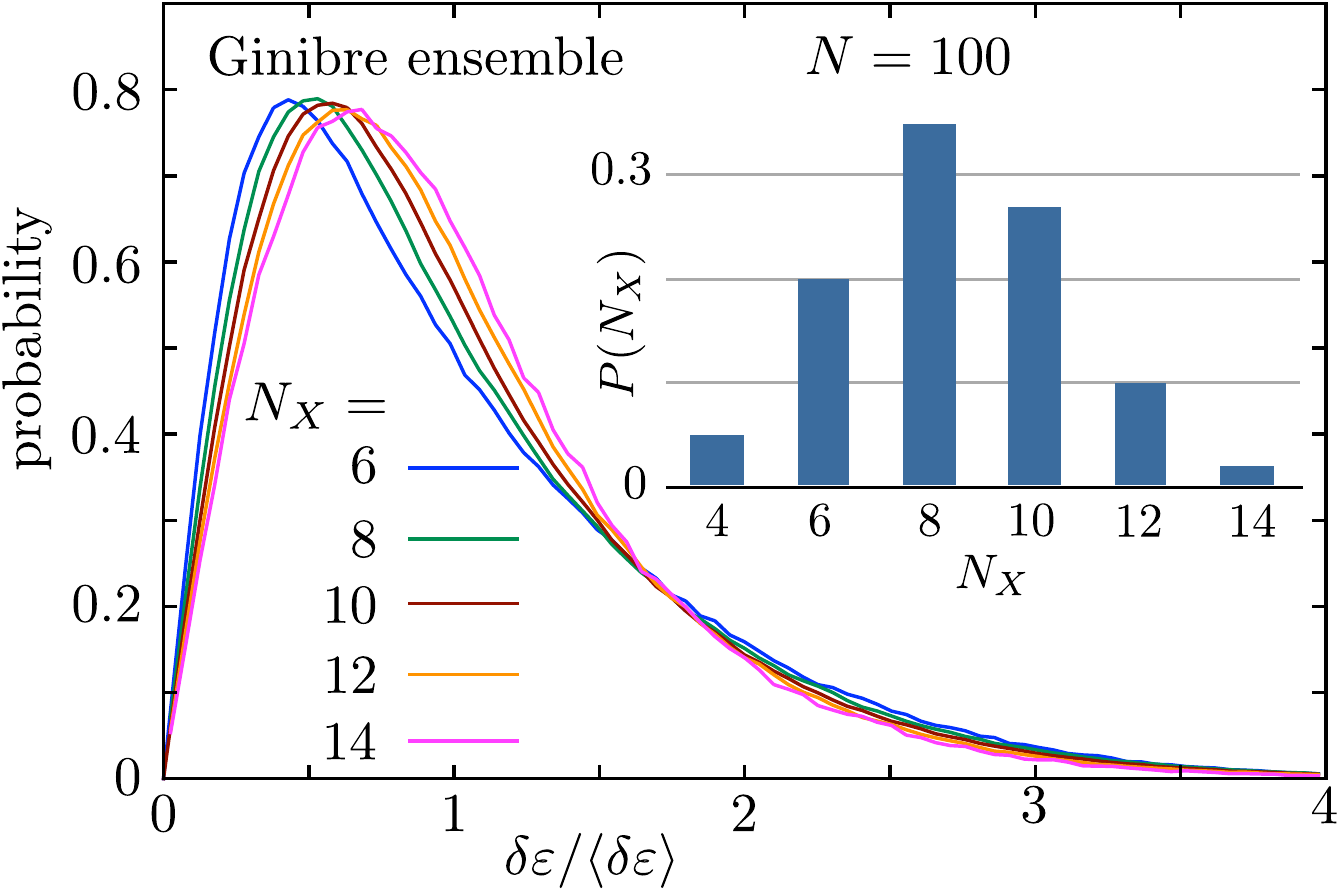}}
\caption{Distribution of the normalized spacings $s=\delta\varepsilon/\langle\delta\varepsilon\rangle$ of neighboring eigenvalues on the real axis in the Ginibre ensemble of $N\times N$ real Gaussian matrices, with $N=100$. The curves are calculated by generating a large number of random matrices in this ensemble, and separating them in sets with the same number $N_{X}$ of real eigenvalues at average spacing $\langle\delta\varepsilon\rangle\approx 2\sqrt{N}/N_{X}$. Each set has its own spacing distribution $P(s|N_{X})$, shown by the colored curves. The inset shows the fraction $P(N_X)$ of matrices with a given $N_X$.
}
\label{fig_spacingGinibre}
\end{figure}

\begin{figure}[tb]
\centerline{\includegraphics[width=0.9\linewidth]{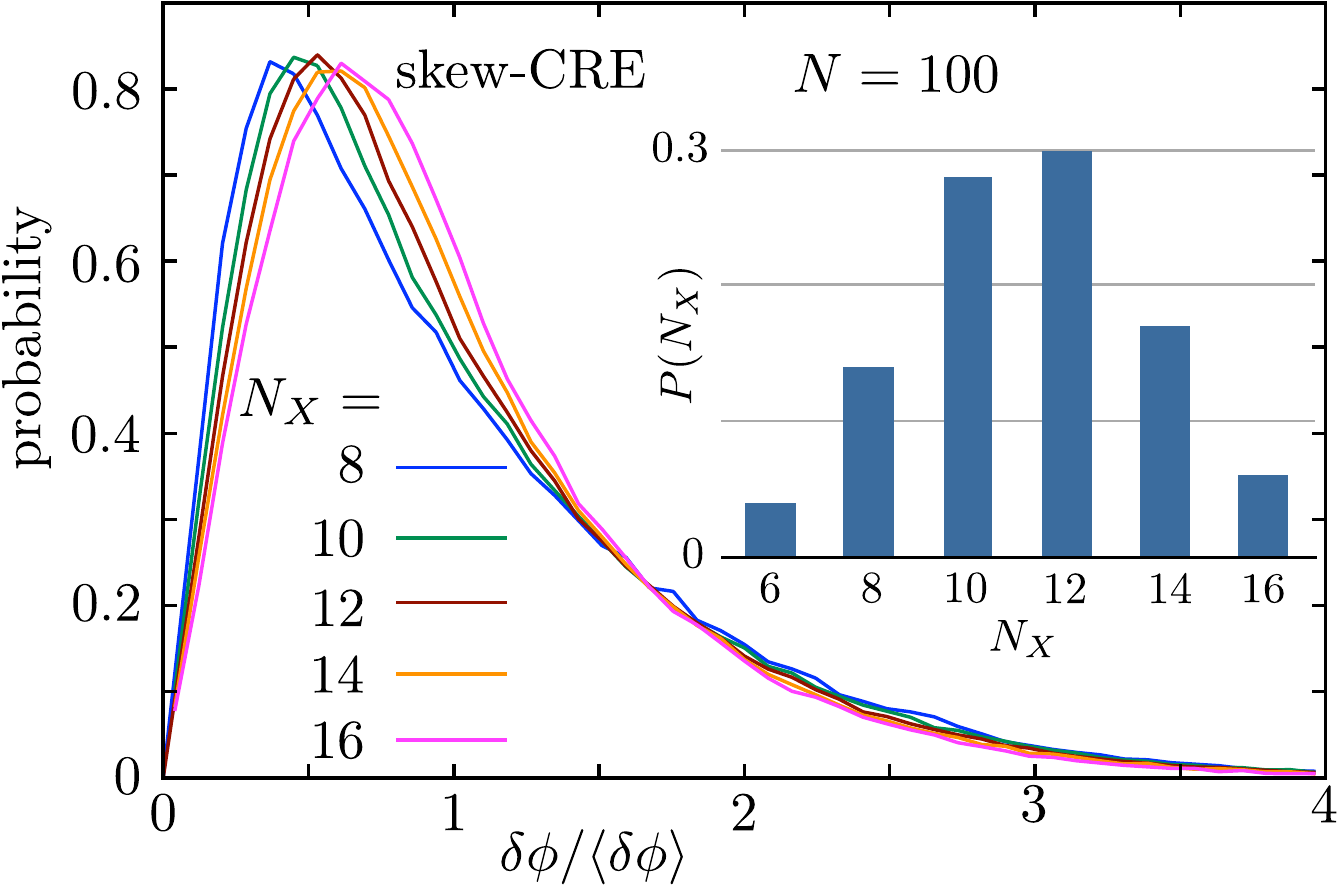}}
\caption{Same as Fig.\ \ref{fig_spacingGinibre}, but now for the ensemble of $2N\times 2N$ skew-Hamiltonian matrices constructed from the Haar measure on ${\rm SO}(2N)$ (skew-CRE). To achieve a uniform spacing, the $N_{X}$ distinct real $\varepsilon$'s are mapped onto the unit circle by $\phi=2\,{\rm arctan}\,\varepsilon$, with average spacing $\langle\delta\phi\rangle=2\pi/N_{X}$.
}
\label{fig_spacingCUE}
\end{figure}

In the Gaussian orthogonal ensemble (GOE) of real symmetric matrices the spacing $\delta \varepsilon=|\varepsilon_{n+1}-\varepsilon_{n}|$ of subsequent eigenvalues $\varepsilon_{n}$ is well described by the Wigner surmise \cite{Meh04,For10},
\begin{equation}
P_{\rm Wigner}(s)=\tfrac{1}{2}\pi s \exp\bigl(-\tfrac{1}{4}\pi s^{2}\bigr),\;\;s=\delta \varepsilon/\langle\delta \varepsilon\rangle.\label{Wigner}
\end{equation}
The spacing distribution vanishes as $s^{\beta}$ with $\beta=1$ for small spacings, a characteristic feature of the GOE known as linear level repulsion.

Linear repulsion applies as well to the Ginibre ensemble of $N\times N$ Gaussian matrices without any symmetry \cite{Leh91,For07,Kho11}. As shown in Fig.\ \ref{fig_spacingGinibre}, the linear repulsion $P(s|N_{X})=C_{N_{X}}s+{\cal O}(s^{2})$ is universal but the slope $C_{N_{X}}$ depends on the number $N_{X}$ of real eigenvalues: smaller $N_{X}$ gives a larger slope. We interpret this is as a ``screening'' effect of nearby complex eigenvalues, which soften the repulsion of neigboring real eigenvalues. Since the average spacing $\langle\delta\varepsilon\rangle\approx 2\sqrt{N}/N_{X}$ of the real eigenvalues is larger for smaller $N_{X}$, there are more intermediate complex eigenvalues for smaller $N_{X}$, consistent with the weaker repulsion.\footnote{J. Bloch, F. Bruckmann, N. Meyer, and S. Schierenberg, JHEP \textbf{08}, 066 (2012),
argue for a \textit{stronger} repulsion of real eigenvalues due to intermediate complex eigenvalues, which is not what we find.}

The Ginibre ensemble has an approximately uniform eigenvalue density on the real axis, while the skew-Hamiltonian ensembles derived from the CUE or CRE have the strongly nonuniform density \eqref{rhoepsilon}. To calculate the spacing distribution in those ensembles (skew-CUE and skew-CRE) we map the real axis onto the unit circle via the transformation \eqref{epsilonzmapping}. The phases $\phi_{n}=2\,{\rm arctan}\,\varepsilon_{n}$ on the unit circle have a uniform density, so for each number $N_{X}$ of distinct real eigenvalues there is a uniform average spacing $\langle\delta\phi\rangle=2\pi/N_{X}$. The spacing distributions $P(s|{N_{X}})$ are very similar to those in the Ginibre ensemble, see Fig.\ \ref{fig_spacingCUE}.

\begin{figure}[tb]
\centerline{\includegraphics[width=0.9\linewidth]{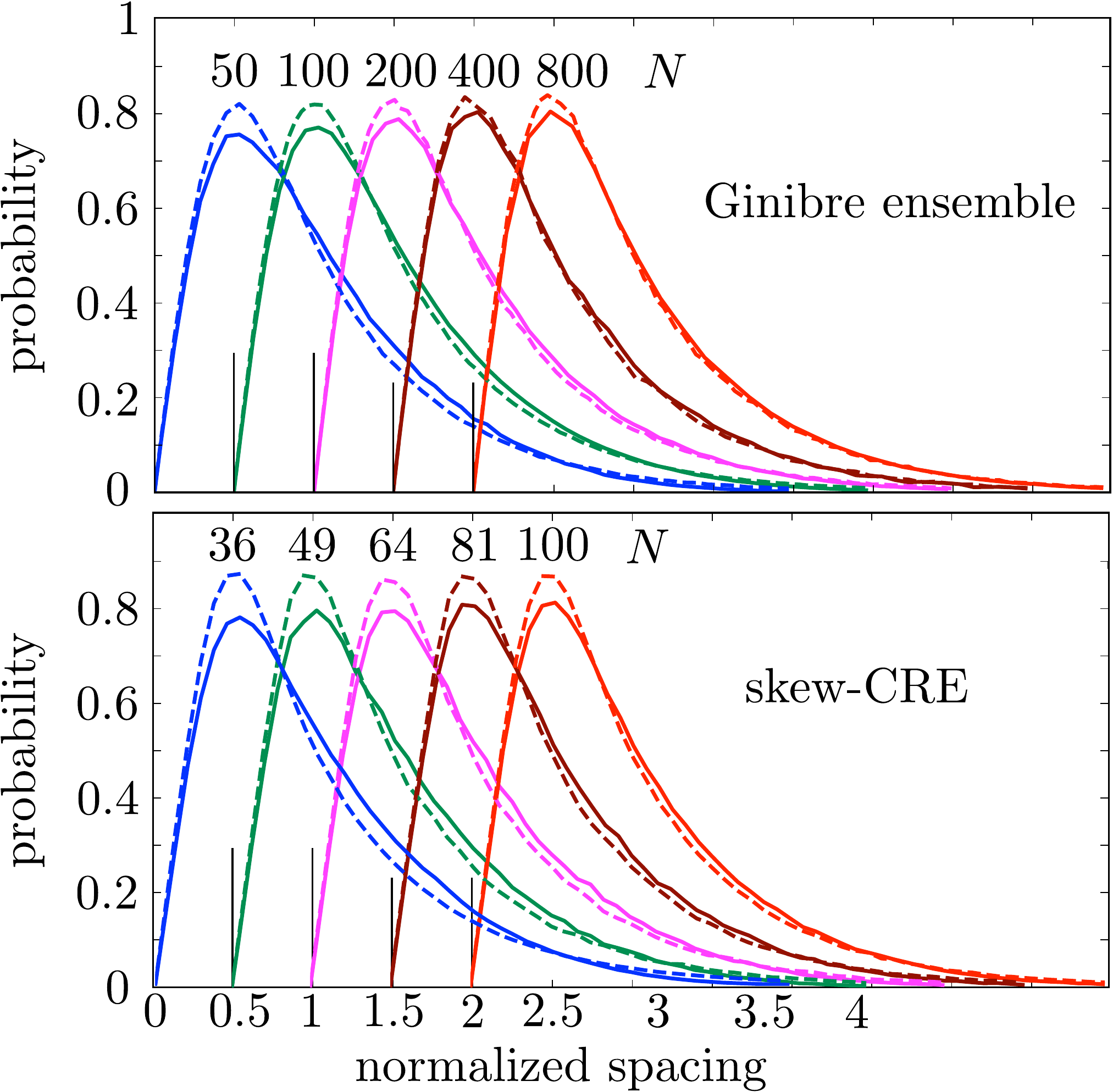}}
\caption{Spacing distributions of real eigenvalues in the Ginibre ensemble ($N\times N$ real Gaussian matrices) and the skew-CRE [$2N\times 2N$ skew-Hamiltonian matrices constructed from ${\rm SO}(2N)$], for different matrix size $N$. All curves cover the interval $0<s<4$, with a different horizontal offset for each $N$. The solid curves are the cumulative average \eqref{Pc} over $N_{X}$-dependent spacing distributions and the dashed curves are a global average over the entire ensemble. The difference amounts to a different way of normalizing by the average spacing. For the cumulative average we normalize with an $N_{X}$-dependent average spacing, for the global average we have a single average spacing for the entire ensemble.
}
\label{fig_spacing_cumulative}
\end{figure}

\begin{figure}[tb]
\centerline{\includegraphics[width=0.9\linewidth]{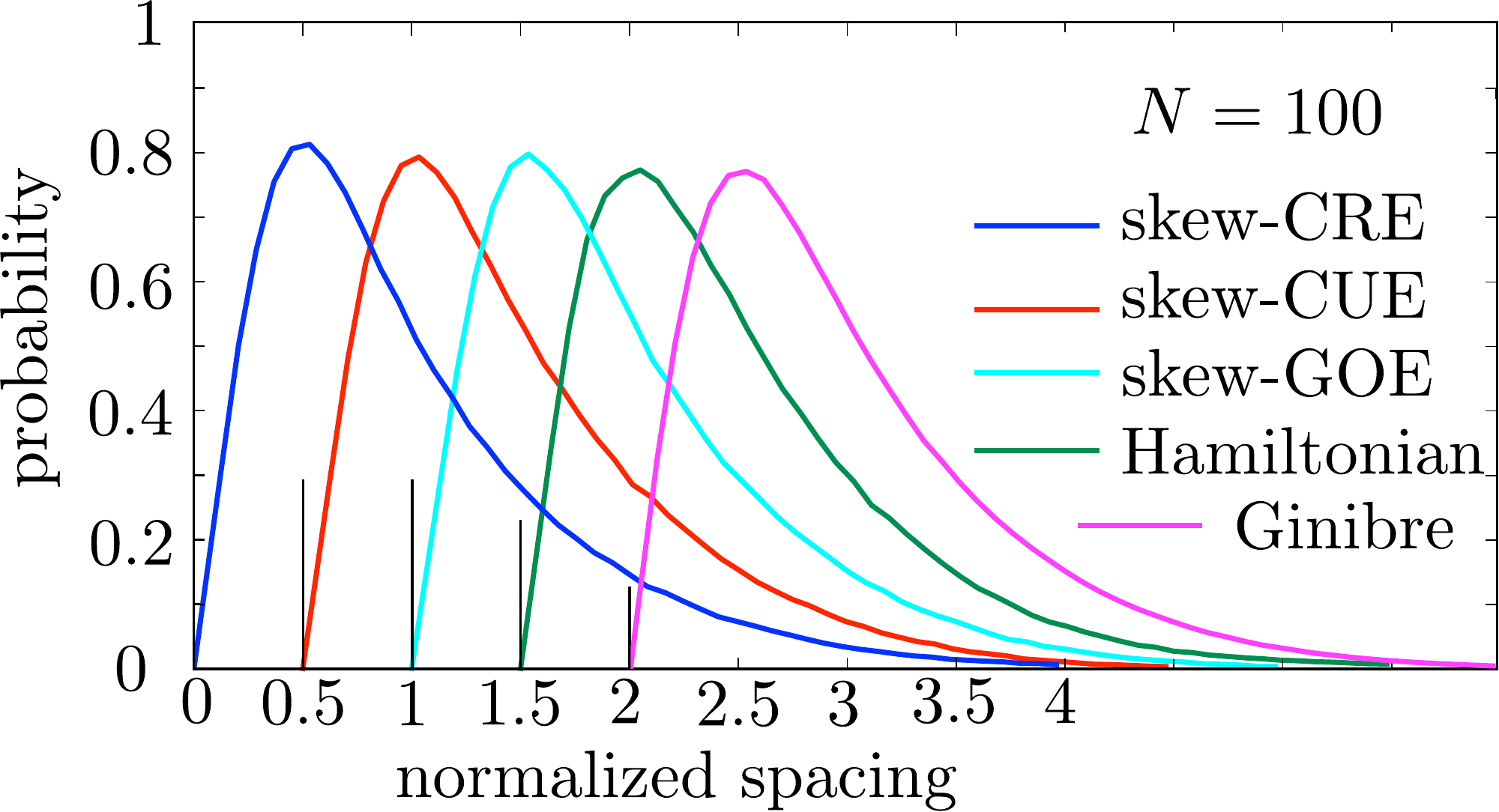}}
\caption{Comparison of the cumulative spacing distribution of real eigenvalues in the five different ensembles of real random matrices. For clarity, each curve has a different horizontal offset.
}
\label{fig_spacing_comparison}
\end{figure}

From $P(s|N_{X})$ we can calculate the cumulative spacing distribution,
\begin{equation}
P_{\rm cumul}(s)=\sum_{N_{X}}P(s|N_{X})P(N_{X}),\label{Pc}
\end{equation}
as the weighted average over different values of $N_{X}$. Since the average spacing is $N_{X}$-dependent, the cumulative spacing distribution \eqref{Pc} is different from the global spacing distribution $P_{\rm global}$, obtained by normalizing the spacing by the average spacing of real eigenvalues in the entire ensemble. Both distributions are shown in Fig.\ \ref{fig_spacing_cumulative}, and one sees that the difference is small. (We do not know whether $P_{\rm cumul}$ and $P_{\rm global}$ become identical in the large-$N$ limit.) Fig.\ \ref{fig_spacing_comparison} shows that the difference between one ensemble and the other is also small, indicating that these random-matrix ensembles have a universal spacing distribution of real eigenvalues.

\begin{figure}[tb]
\centerline{\includegraphics[width=0.9\linewidth]{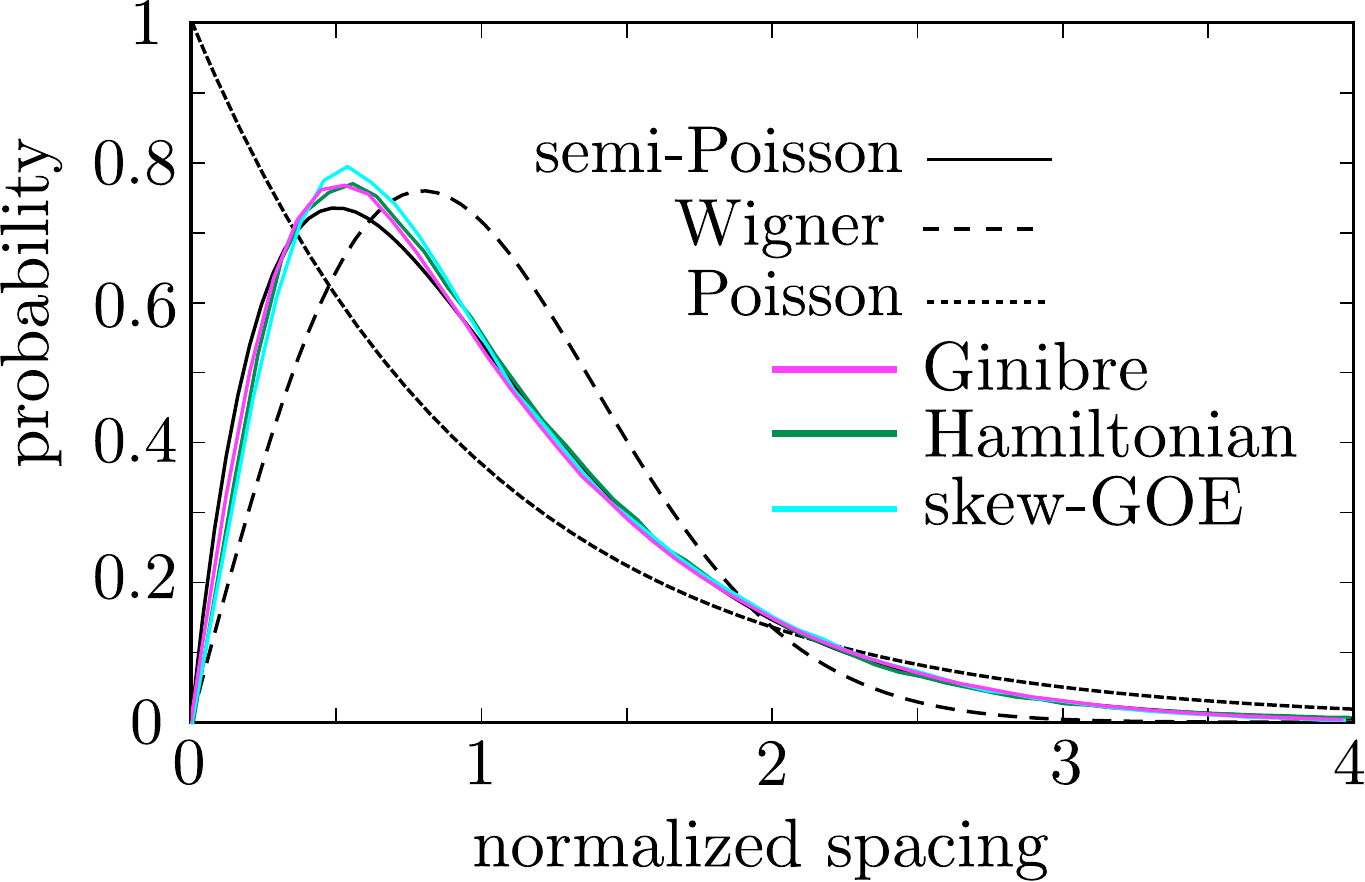}}
\caption{Solid curves: cumulative spacing distribution on the real axis for $N=100$ in the three non-Hermitian ensembles with Gaussian measure. The corresponding figure for the two ensembles with Haar measure is Fig.\ \ref{fig_semiPoisson} in the main text. Together these figures show that the spacing distribution on the real axis is close to the semi-Poissonian form.
}
\label{fig_semiPoisson_Ginibre}
\end{figure}

\begin{figure}[tb]
\centerline{\includegraphics[width=0.9\linewidth]{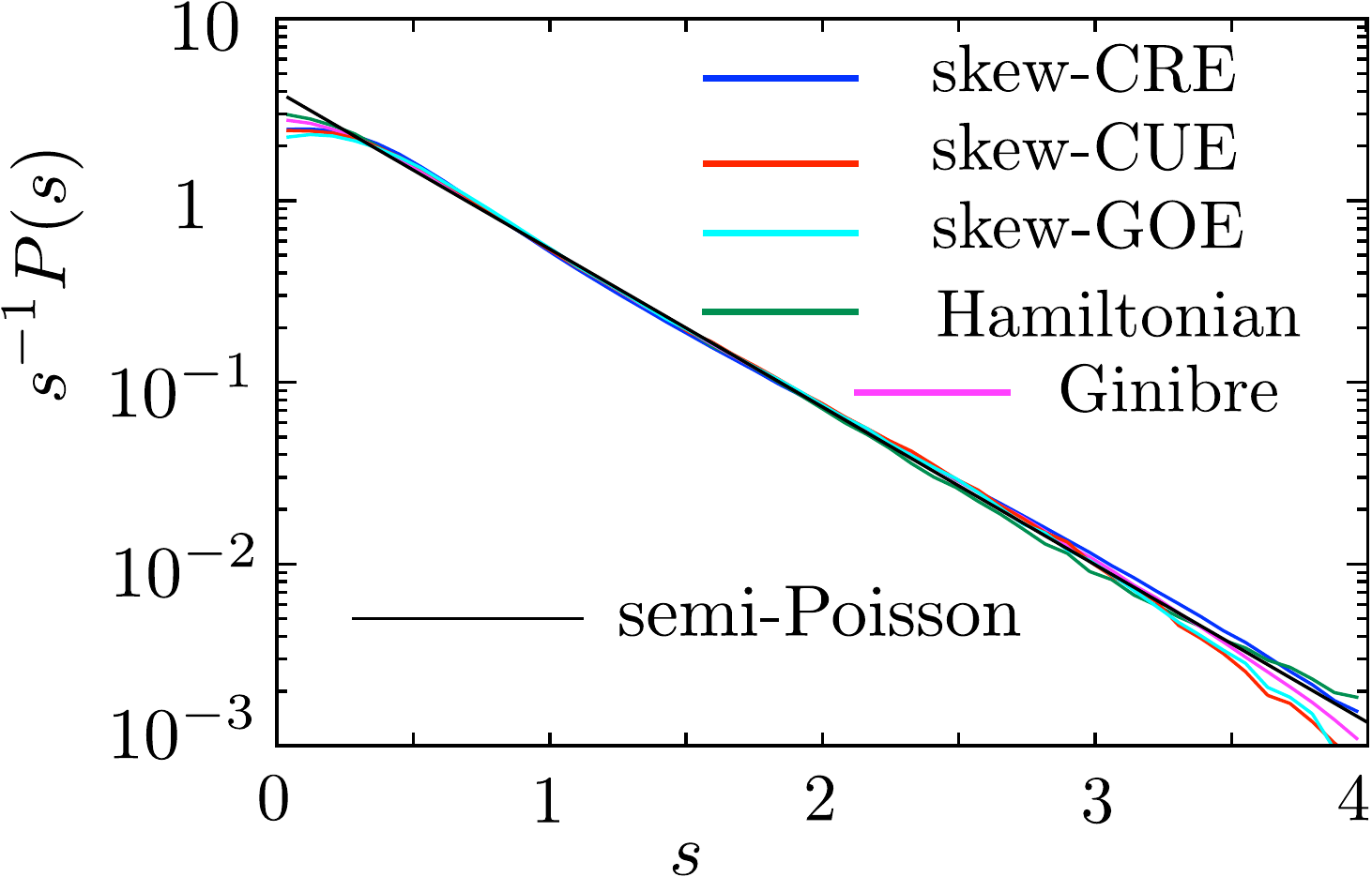}}
\caption{Cumulative spacing distribution on the real axis for all five ensembles (with $N=100$), plotted on a log-linear scale to demonstrate the semi-Poissonian tail.
}
\label{fig_logplot}
\end{figure}

In Fig.\ \ref{fig_semiPoisson} in the main text the cumulative spacing distributions on the real axis in the skew-CRE and skew-CUE are compared with the Wigner distribution \eqref{Wigner}. A similar comparison for the skew-GOE, Hamiltonian, and Ginibre ensembles is shown in Fig.\ \ref{fig_semiPoisson_Ginibre}. We also compare with the Poisson distribution of uncorrelated eigenvalues,
\begin{equation}
P_{\rm Poisson}(s)=e^{-s},\label{Poisson}
\end{equation}
and with the semi-Poisson distribution \cite{Bog99},
\begin{equation}
P_{\text{semi-Poisson}}=4se^{-2s}.\label{semiPoisson}
\end{equation}
None of these three distributions fits the RMT data precisely, but the semi-Poissonian form describes most closely both the linear repulsion at small spacings and the exponential tail at large spacings (see also Fig.\ \ref{fig_logplot} for a log-linear plot).

\begin{figure}[tb]
\centerline{\includegraphics[width=0.9\linewidth]{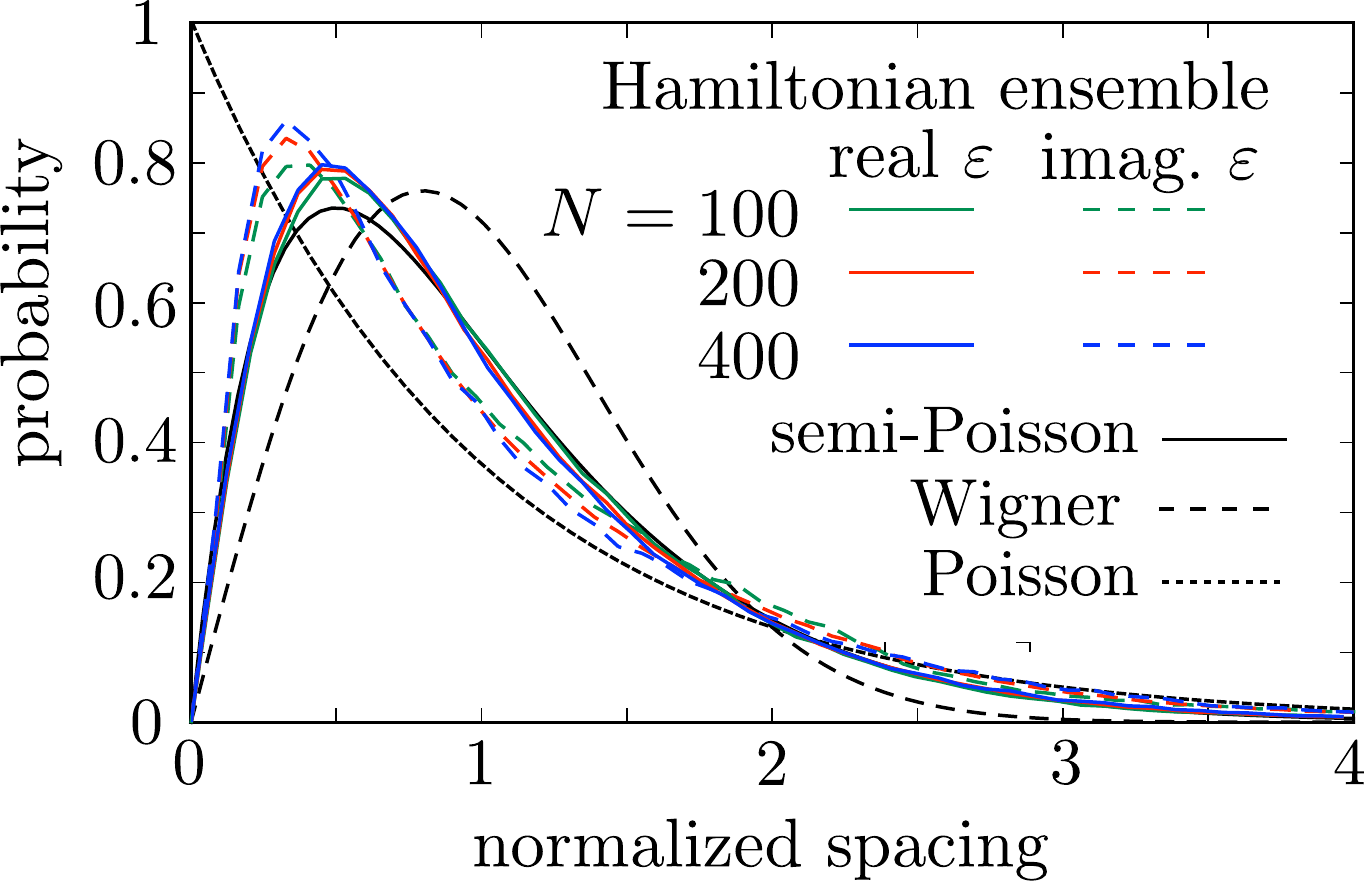}}
\caption{Cumulative spacing distribution on the real axis (solid colored curves) and on the imaginary axis (dashed colored curves) for different values of $N$ in the Hamiltonian ensemble.
}
\label{fig_semiPoisson_Hamiltonian}
\end{figure}

In the ensemble of Hamiltonian matrices it is of interest to compare the spacing distributions on the real and on the imaginary axis, see Fig.\ \ref{fig_semiPoisson_Hamiltonian}. The spacing distributions are qualitatively similar, but distinct --- we show different values of $N$ to confirm that the difference is not a finite-size effect.

These are all numerical findings. In the Ginibre ensemble the complete joint probability distribution of the eigenvalues is known analytically \cite{Leh91,For07}. It might be feasible to derive the spacing distribution on the real axis from that joint distribution and see how close it approaches the semi-Poissonian form in the large-$N$ limit. Such a calculation might also confirm our intuition that a screening effect of complex eigenvalues is responsible for the Wigner-to-Poisson crossover with increasing spacing.


\begin{thebibliography}{99}
\bibitem{Neu29} J. von Neumann and E. Wigner, Phys. Z. \textbf{30}, 467 (1929).
\bibitem{Alt97} A. Altland and M. R. Zirnbauer, Phys. Rev. B \textbf{55}, 1142 (1997).
\bibitem{Kit01} A. Yu. Kitaev, Phys. Usp. \textbf{44} (suppl.), 131 (2001).
\bibitem{Sak70} A. Sakurai, Prog. Theor. Phys. \textbf{44}, 1472 (1970).
\bibitem{Bal06} A. V. Balatsky, I. Vekhter, and J.-X. Zhu, Rev. Mod. Phys. \textbf{78}, 373 (2006).
\bibitem{Ryu10} S. Ryu, A. Schnyder, A. Furusaki, and A. Ludwig, New J. Phys. \textbf{12}, 065010 (2010).
\bibitem{And11} B. M. Andersen, K. Flensberg, V. Koerting, and J. Paaske, Phys. Rev. Lett. \textbf{107}, 256802 (2011).
\bibitem{Law11} K. T. Law and P. A. Lee, Phys. Rev. B \textbf{84}, 081304 (2011).
\bibitem{Bee13} C. W. J. Beenakker, D. I. Pikulin, T. Hyart, H. Schomerus, and J. P. Dahlhaus, Phys. Rev. Lett. \textbf{110}, 017003 (2013).
\bibitem{Yok13} T. Yokoyama, M. Eto, and Yu. V. Nazarov, J. Phys. Soc. Jpn. \textbf{82}, 054703 (2013).
\bibitem{Lee12} E. J. H. Lee, X. Jiang, R. Aguado, G. Katsaros, C. M. Lieber, and S. De Franceschi, Phys. Rev. Lett. \textbf{109}, 186802 (2012); E. J. H. Lee, \textit{et al.}, arXiv:1302.2611.
\bibitem{Cha12} W. Chang, V. E. Manucharyan, T. S. Jespersen, J. Nyg{\aa}rd, and C. M. Marcus, Phys. Rev. Lett. \textbf{110}, 217005 (2013).
\bibitem{Sau12} J. D. Sau and E. Demler, arXiv:1204.2537.
\bibitem{Ila13} R. Ilan, J. H. Bardarson, H.-S. Sim, and J. E. Moore, arXiv:1305.2210.
\bibitem{Kwo04} H.-J. Kwon, K. Sengupta, and V. M. Yakovenko, Eur. Phys. J. B \textbf{37}, 349 (2004).
\bibitem{Mi13}  S. Mi, D. I. Pikulin, M. Wimmer, and C. W. J. Beenakker, Phys. Rev. B \textbf{87}, 241405(R) (2013).
\bibitem{Meh04} M. L. Mehta, \textit{Random Matrices} (Elsevier, 2004).
\bibitem{For10} P. J. Forrester, \textit{Log-Gases and Random Matrices} (Princeton University Press, 2010).
\bibitem{Shk93} B. I. Shklovskii, B. Shapiro, B. R. Sears, P. Lambrianides, and H. B. Shore, Phys. Rev. B \textbf{47}, 11487 (1993).
\bibitem{Bog99} E. B. Bogomolny, U. Gerland, and C. Schmit, Phys. Rev. E \textbf{59}, R1315 (1999).
\bibitem{Bee05} C. W. J. Beenakker, Lect. Notes Phys. \textbf{667}, 131 (2005) [arXiv:cond-mat/0406018].
\bibitem{Ali12} J. Alicea, Rep. Prog. Phys. \textbf{75}, 076501 (2012) [arXiv:1202.1293].
\bibitem{Akh11} A. R. Akhmerov, J. P. Dahlhaus, F. Hassler, M. Wimmer, and C. W. J. Beenakker, Phys. Rev. Lett. \textbf{106}, 057001 (2011).
\bibitem{Bee11} C. W. J. Beenakker, J. P. Dahlhaus, M. Wimmer, and A. R. Akhmerov, Phys. Rev. B \textbf{83}, 085413 (2011).
\bibitem{note1} The Cayley transform \eqref{AOrelation} does not exist if ${\rm Det}\,O=-1$, which happens if one superconductor is topologically trivial and the other nontrivial. Then the quantum dot contains an unpaired Majorana zero mode at any value of $\phi$, so the whole notion of level crossings loses its meaning.
\bibitem{note2} The twofold degeneracy of the eigenvalues of a skew-Hamiltonian matrix follows directly from the fact that  the determinant \eqref{DetA} can equivalently be written as the square of a Pfaffian: $[{\rm Pf}\,(A+\varepsilon J)]^{2}=0$. The sign of this Pfaffian is the topological quantum number representing the ground-state fermion parity.
\bibitem{Leh91} N. Lehmann and H.-J. Sommers, Phys. Rev. Lett. \textbf{67}, 941 (1991).
\bibitem{Ede94} A. Edelman, E. Kostlan, and M. Shub, J. Am. Math. Soc. \textbf{7}, 247 (1994); A. Edelman, J. Multivariate Anal. \textbf{60}, 203 (1997).
\bibitem{Kan05} E. Kanzieper and G. Akemann, Phys. Rev. Lett. \textbf{95}, 230201 (2005).
\bibitem{For07} P. J. Forrester and T. Nagao, Phys. Rev. Lett. \textbf{99}, 050603 (2007).
\bibitem{Kho11} B. A. Khoruzhenko and H.-J. Sommers, arXiv:0911.5645, published in: \textit{Handbook on Random Matrix Theory}, edited by G. Akemann, J. Baik, and P. Di Francesco (Oxford University Press, Oxford, 2011).
\bibitem{Gin65} J. Ginibre, J. Math. Phys. \textbf{6}, 440 (1965).
\bibitem{Gir85} V. L. Girko, Theory Prob. Appl. \textbf{29}, 694 (1985).
\bibitem{Tao10} T. Tao and V. Vu, Ann. Probab. \textbf{38}, 2023 (2010).
\bibitem{Bor12} C. Bordenave and D. Chafa\"{i}, Prob. Surveys \textbf{9}, 1 (2012) [arXiv:1109.3343].
\bibitem{RMT_app} Random unitary and orthogonal matrices were generated using the method of F. Mezzadri, Notices Am. Math. Soc. \textbf{54}, 592 (2007). Eigenvalues of skew-Hamiltonian matrices were calculated using the algorithm of P. Benner, D. Kressner, and V. Mehrmann, in: \textit{Proceedings of the Conference on Applied Mathematics and Scientific Computing} (Springer, Amsterdam, 2005). For details of these random-matrix calculations we refer to App.\ \ref{RMTdetails}.
\bibitem{Gor01} T. Gorin, M. M\"{u}ller, and P. Seba, Phys. Rev. E \textbf{63}, 068201 (2001).
\bibitem{Gar06} A. M. Garc\'{\i}a-Garc\'{\i}a and J. Wang, Phys. Rev. E \textbf{73}, 036210 (2006).
\bibitem{Dys62} F. J. Dyson, J. Math. Phys. \textbf{3}, 140 (1962).
\bibitem{note3} We have found that the same hybrid Wigner-Poisson spacing distribution of real eigenvalues applies also to the Ginibre ensemble, see  App.\ \ref{RMTdetails}.
\bibitem{Rok12} L. P. Rokhinson, X. Liu, and J. K. Furdyna, Nature Phys. \textbf{8}, 795 (2012).
\bibitem{kwant} The tight-binding model calculations were performed using the {\sc kwant} package developed by A. R. Akhmerov, C. W. Groth, X. Waintal, and M. Wimmer, {\tt www.kwant-project.org}. To efficiently calculate the lowest energy levels we used the {\sc arpack} package developed by R. Lehoucq, K. Maschhoff, D. Sorensen, and C. Yang, {\tt www.caam.rice.edu/software/ARPACK}.
\bibitem{InSbparam} Parameters used for the InSb model calculation in Fig.\ \ref{fig_wire} are: magnetic field $B=6.2\,{\rm mT}$, corresponding to a flux $3h/e$ through an area $L\times W=2\,\mu{\rm m}\times 1\,\mu{\rm m}$, spin-orbit length $l_{\rm so}=0.25\,\mu{\rm m}$, lattice constant $a=l_{\rm so}/10$, Fermi energy $E_{\rm F}=2.5\,{\rm meV}$, corresponding to $N=20$ transverse modes, superconducting gap $\Delta_{0}=0.4\,{\rm meV}$, and disorder strength $U_{0}=1.7\,{\rm meV}$, corresponding to a mean free path $l\approx L/2$.
\bibitem{Liu08} C. Liu, T. L. Hughes, X.-L. Qi, K. Wang, and S.-C. Zhang, Phys. Rev. Lett. \textbf{100}, 236601 (2008).
\bibitem{Kne12} I. Knez, R.-R. Du, and G. Sullivan, Phys. Rev. Lett. \textbf{107}, 136603 (2011); \textbf{109}, 186603 (2012).
\bibitem{Ber06} B. A. Bernevig, T. L. Hughes, and S. C. Zhang, Science \textbf{314}, 1757 (2006).
\bibitem{Has10} M. Z. Hasan and C. L. Kane, Rev. Mod. Phys. \textbf{82}, 3045 (2010)
\bibitem{Qi11} X.-L. Qi and S.-C. Zhang, Rev. Mod. Phys. \textbf{83}, 1057 (2011).

\end{thebibliography}
\end{document}